\documentclass[12pt]{iopart}
\usepackage{xcolor}
\usepackage{hyperref}
\usepackage{macros}
\usepackage{multirow}
\usepackage{amsmath}
\usepackage{amsfonts}
\usepackage{dsfont}
\usepackage{tcolorbox}
\usepackage{graphicx}
\usepackage{etoolbox}

\makeatletter
\def\@mkboth#1#2{}
\newlength\appendixwidth
\preto\appendix{\addtocontents{toc}{\protect\patchl@section}}
\newcommand{\patchl@section}{%
  \settowidth{\appendixwidth}{\textbf{Appendix }}%
  \addtolength{\appendixwidth}{1.5em}%
  \patchcmd{\l@section}{1.5em}{\appendixwidth}{}{\ddt}%
}
\makeatother

\begin{document}

\review[Scars and Fragmentation: Exact Results]{Quantum Many-Body Scars and Hilbert Space Fragmentation: A Review of Exact Results}

\author{Sanjay Moudgalya,$^{1,2}$ B. Andrei Bernevig$^{3,4,5}$ and Nicolas Regnault$^{3,6}$}

\address{$^{1}$ Institute for Quantum Information and Matter, California Institute of Technology, Pasadena, CA 91125, USA\\
$^{2}$ Walter Burke Institute for Theoretical Physics, California Institute of Technology, Pasadena, California 91125, USA\\
$^{3}$ Department of Physics, Princeton University, Princeton, NJ 08544, USA\\
$^{4}$ Donostia International Physics Center, P. Manuel de Lardizabal 4, 20018 Donostia-San Sebastian, Spain\\
$^{5}$ IKERBASQUE, Basque Foundation for Science, Bilbao, Spain\\
$^{6}$ Laboratoire de Physique de l'Ecole normale sup\'{e}rieure, ENS, Universit\'{e} PSL, CNRS, Sorbonne Universit\'{e}, Universit\'{e} Paris-Diderot, Sorbonne Paris Cit\'{e}, 75005 Paris, France}

\begin{abstract}
The discovery of Quantum Many-Body Scars (QMBS) both in Rydberg atom simulators and in the Affleck-Kennedy-Lieb-Tasaki (AKLT) spin-1 chain model, have shown that a weak violation of ergodicity can still lead to rich experimental and theoretical physics.
In this review, we provide a pedagogical introduction to and an overview of the exact results on weak ergodicity breaking via QMBS in isolated quantum systems with the help of simple examples such as the fermionic Hubbard model.
We also discuss various mechanisms and unifying formalisms that have been proposed to encompass the plethora of systems exhibiting QMBS.
We cover examples of equally-spaced towers that lead to exact revivals for particular initial states, as well as isolated examples of QMBS.
Finally, we review Hilbert Space Fragmentation, a related phenomenon where systems exhibit a richer variety of ergodic and non-ergodic behaviors, and discuss its connections to QMBS.
\end{abstract}
\tableofcontents
\section{Introduction}\label{sec:intro}
\begin{figure}
    \centering
    \begin{tabular}{c}
    \includegraphics[scale=0.8]{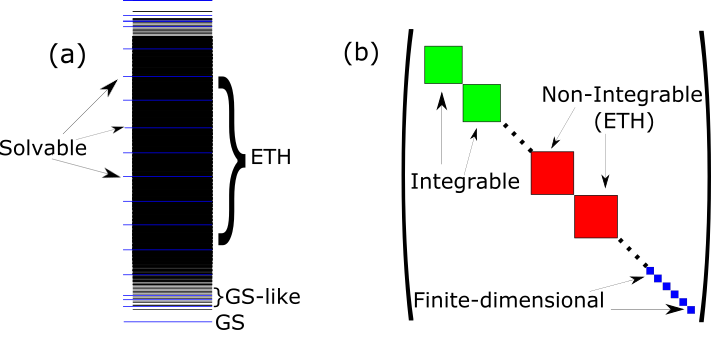}
    \end{tabular}
    \caption{(Color online) Two types of ergodicity breaking that we study in this review. (a) Quantum Many-Body Scars (QMBS): typical example of an energy spectrum with solvable ETH-violating eigenstates that show sub-volume law entanglement and exist amidst a sea of ETH-satisfying states that show volume-law entanglement. States close to the edges of the spectrum such as the ground state (GS) or low-energy excitations (GS-like) show area-law entanglement and are not expected to satisfy ETH. (b) Hilbert Space Fragmentation: Hamiltonian matrix represented consists of several dynamically disconnected Krylov subspaces, i.e. it is block-diagonal in a certain simple basis (e.g., the product state basis). The size of the Krylov subspaces can grow with system size or stay finite-dimensional, and the Hamiltonian restricted to the Krylov subspaces can be integrable (green) or non-integrable (red). The Hamiltonian in non-integrable Krylov subspaces is expected to satisfy Krylov-Restricted ETH.}
    \label{fig:main}
\end{figure}

The advent of quantum simulators, implemented for example in ultracold atomic setups or superconducting circuits, has put at the forefront the question of out-of-equilibrium quantum many-body systems.
The Eigenstate Thermalization Hypothesis (ETH)~\cite{deutsch1991quantum,srednicki1994chaos} has long been thought to describe the properties of \textit{all} finite-energy density eigenstates, i.e., excited states in the middle of the energy spectrum, of any generic non-integrable quantum many-body system.
ETH has been tested experimentally, analytically and numerically in various systems (however, mostly in one dimension), and it forms the pillar of our understanding of phenomena such as many-body quantum chaos and thermalization. 
While the formulation of ETH sounds general, it is nevertheless a hypothesis, and thus immediately raises the question of potential counter-examples. 
Among them, quantum integrable models are the simplest cases where violations of ETH are known.
Such systems exhibit an extensive number of conserved quantities that in principle determine every eigenstate in the system, which exhibit features that might strongly deviate from a typical thermal eigenstates.
While non-interacting integrable models such as free fermion models are ubiquitous in physics, interacting ones are usually considered as fine-tuned and are harder to experimentally implement.
The search for more generic violations of ETH beyond integrable systems began with the discovery of many-body localization (MBL), where the alliance of strong disorder and interaction leads to emergent integrability (see for example the two review articles Refs.~\cite{nandkishore2015many} and~\cite{abanin2018review}), although the existence and stability of MBL is currently being debated~\cite{suntajs2020quantum, abanin2021distinguishing}. 
The intermediate situation, a weak or partial violation of ETH by a small number (exponentially smaller than the Hilbert space dimension) of eigenstates, might be considered at first sight as too non-generic to be interesting or experimentally relevant.
Indeed, any such non-thermal eigenstate would not have an energy gap protecting its nature, and would be exponentially close in energy to thermal eigenstates that would quickly hybridize with it under small perturbations.
However, an experimental observation of anomalously long-lived revivals in a Rydberg atom quantum simulator~\cite{bernien2017probing} showed the opposite; the revivals were attributed to a small set of non-thermal eigenstates, dubbed Quantum Many-Body Scars (QMBS), in the otherwise non-integrable PXP model that captured the experiment~\cite{turner2017quantum, turner2018quantum}. 
These results on Rydberg atoms also led to numerous further theoretical investigations of the PXP model~\cite{ho2018periodic, khemani2019int, choi2018emergent}, aspects of which have been summarized in the recent review Ref.~\cite{SerbynReview} (see Ref.~\cite{papic2021weak} for a longer version).
The typical spectrum of a system exhibiting QMBS is depicted in Fig.~\ref{fig:main}a: a discrete number of non-ETH eigenstates that ``scar" the spectrum of an apparently ergodic system.
With a proper choice of an experimentally motivated initial state, the time-evolution of the quantum many-body system would then show a strong departure from the typical behavior of thermal non-integrable models.
In parallel to this experimental breakthrough, QMBS were independently discovered using a purely theoretical approach in a different context.
Ref.~\cite{moudgalya2018a} derived in the Affleck-Kennedy-Lieb-Tasaki (AKLT) spin-1  model~\cite{aklt1987rigorous} a series of energetically equally-spaced exact excited eigenstates, i.e., a tower of states, that provably violates ETH in an otherwise non-integrable model.
This led to a flurry of analytical results that provided a complementary perspective on QMBS and attempted to establish a rigorous understanding of its emergence based on either brute force analytical derivation of excited states, or an underlying algebraic structure or symmetry.
This should be put in contrast with some other forms of ergodicity breaking, e.g., MBL, where analytical progress has been hindered by the scarcity of exact results.
Beyond the AKLT model, QMBS have been found in a variety of systems, sometime giving the opportunity to revisit some of the most celebrated condensed-matter interacting models such as the Hubbard model, in the search for analytical expression of exact excited states.
QMBS are also closely related to the broader phenomenon of Hilbert space fragmentation~\cite{sala2019ergodicity} (also referred to as Hilbert space shattering~\cite{khemani2019local}, Krylov fracture~\cite{moudgalya2019thermalization}, or jamming~\cite{zadnik2021foldedxxz1}), which refers to the existence of exponentially many dynamically disconnected subspaces that are not captured by conventional symmetries. 
There, a physically motivated basis choice unveils a rich structure within the Hamiltonian of dynamically disconnected subspaces with different thermalization and entanglement properties, as sketched in Fig.~\ref{fig:main}b.
Akin to QMBS, analytically tractable models offer an invaluable playground to understand the nature of Hilbert space fragmentation and their effects on dynamics.
In this review, we focus on exact analytical results about QMBS which already provide a wealth of interesting models, analytical derivations and formalisms.
We refer the readers leaning towards direct experimental implications of QMBS or in the approximate QMBS of PXP and related models to Ref.~\cite{SerbynReview}.
The review is organized as follows.
In Sec.~\ref{sec:ergodicity}, we provide, for pedagogical purposes, a short overview on ergodicity and its breakdown in isolated quantum systems, introducing notations and concepts used in QMBS literature.
Sec.~\ref{sec:towers} focuses on towers of QMBS and explicitly illustrates the towers derived from the spectrum generating algebra or dynamical symmetry in the fermionic Hubbard model.
We also survey other examples of towers of QMBS in the literature, discuss their entanglement properties, and demonstrate how they lead to revivals from simple initial states.
Sec.~\ref{sec:unifiedformalism} gives an overview of the different known mechanisms inducing towers of QMBS, namely the eigenstate embedding, the spectrum generating algebra and their generalizations, and the symmetry-based formalisms.
We dedicate Sec.~\ref{sec:isolated} to reviewing several examples of isolated QMBS, due to its connection to some exact results in the PXP model, as well as a highly general formalism for embedding exact QMBS into the spectrum of any non-integrable Hamiltonian.
In Sec.~\ref{sec:fragmentation} we review ergodicity breaking via the broader phenomenon of Hilbert space fragmentation, exemplified through dipole-conserving systems, and discuss their dynamical implications and connections to QMBS. 
Finally we discuss some major questions still open in the field in Sec.~\ref{sec:outlook}.
\section{Ergodicity in Isolated Quantum Systems}\label{sec:ergodicity}
We begin by reviewing aspects of dynamics of isolated quantum systems. 
For the sake of concreteness,  we focus on a system with $L$ spins and Hamiltonian $H$.
We are typically interested in the dynamics of a simple wavefunction $\ket{\psi(0)}$ under the Hamiltonian $H$, where simple wavefunctions are those that are experimentally accessible,  for example, product states or ground states of simple local Hamiltonians. 
The system evolves the state unitarily,  and the wavefunction of the full system at time $t$ is given by $\ket{\psi(t)} = e^{-i H t}\ket{\psi(0)}$.
An isolated quantum system without any other symmetries is said to be ergodic or thermal if the reduced density matrix of any small subsystem $\ma$ of $L_{\ma} \ll L$ spins,  defined as $\rho_{\ma}(t) \equiv \textrm{Tr}_{\mb}\left(\ket{\psi(t)}\bra{\psi(t)}\right)$ evolves to a Gibbs density matrix.
\begin{equation}
    \lim_{t \rightarrow \infty} \rho_{\ma}(t) = \textrm{Tr}_{\mb}\left(\rho^{eq}\right) \approx \rho^{eq}_{\ma},\;\;\; \rho^{eq} = \frac{1}{Z} e^{-\beta H}
\label{eq:thermalization}
\end{equation}
where $Z$ is the partition function for the subsystem, $\beta$ is an inverse-temperature associated with the initial state.  
In particular, Eq.~(\ref{eq:thermalization}) implies that the rest of the system acts as a thermal bath for the small subsystem $\ma$~\cite{nandkishore2015many}, and as a consequence,  the late-time expectation values of (sums of) local operators $\wO$ that are supported on a small number of sites should match their thermal expectation values. 
In the presence of additional symmetries, Eq.~(\ref{eq:thermalization}) is suitably modified to include a grand canonical ensemble formed by the symmetries, and we refer the readers to detailed reviews on this subject in Refs.~\cite{polkovnikov2011colloquium, d2016quantum}.
These conditions on the dynamics of states have a direct implication on the structure of eigenstates of the system, which we now discuss. 
\subsection{Eigenstate Thermalization Hypothesis (ETH)}\label{subsec:ETH}
The definition of thermalization in Eq.~(\ref{eq:thermalization}) naturally leaves open the question of which initial states $\ket{\psi(0)}$ thermalize. 
Informally speaking, if \emph{any} initial state at some energy density of a system thermalizes under a Hamiltonian $H$, the eigenstates of $H$ at that energy density should also thermalize.
Since we expect such initial state behavior generically, we arrive at the Eigenstate Thermalization Hypothesis (ETH),  which loosely states that any eigenstate of the Hamiltonian at a finite energy density is thermal.
That is, the reduced density matrix of an eigenstate with energy $E_\alpha$ over a small subsystem $\ma$ should also be the Gibbs density matrix over the subsystem with an inverse-temperature $\beta_\alpha$ that depends on $E_\alpha$.
Indeed, we expect $\beta_\alpha \rightarrow \infty$ when $E_\alpha$ is close to the ground state energy, and $\beta_\alpha \rightarrow 0$ when $E_\alpha$ corresponds to the middle of the spectrum.
A more accurate form of ETH is motivated in terms of expectation values of local operators as follows. 
For a system of volume $V$ and a local Hamiltonian $H$, typical initial product states $\ket{\psi(0)}$ have energy variances $\Delta \sim \sqrt{V}$, which, in the thermodynamic limit, is much smaller than the energy bandwidth $W$ which scales as $\sim V$, i.e.~\cite{rigol2008thermalization}
\begin{equation}
    \bra{\psi(0)}H\ket{\psi(0)} = \bar{E},\;\;\; \sqrt{\bra{\psi(0)}H^2\ket{\psi(0)} - \bar{E}^2} = \Delta \ll W.
\label{eq:initstateconditions}
\end{equation}
Hence, when any product state is expressed in the energy eigenbasis $\{\ket{E_\alpha}\}$ of a local Hamiltonian $H$ as $\ket{\psi(0)} = \sumal{\alpha}{}{c_\alpha \ket{E_\alpha}}$, the magnitudes of the coefficients $\{|c_\alpha|^2\}$ turn out to be significant only in an energy window $ E_\alpha \in [\bar{E} - \Delta, \bar{E} + \Delta]$.
The expectation value of a local operator $\wO$ as a function of time then reads
\begin{equation}
    \langle \wO(t) \rangle \equiv \bra{\psi(t)} \wO \ket{\psi(t)} = \sumal{\alpha}{}{|c_\alpha|^2 O_{\alpha\alpha}} + \sumal{\alpha \neq \beta}{}{c_\alpha^\ast c_\beta O_{\alpha\beta} e^{i(E_\alpha - E_\beta)t}},
\label{eq:Otdefn}
\end{equation}
where $O_{\alpha\beta} = \bra{E_\alpha} \wO \ket{E_\beta}$.
Assuming there are no degeneracies in the spectrum, the time-averaged expectation value cancels the off-diagonal terms in Eq.~(\ref{eq:Otdefn}), the long-time average is determined only by the average in the ``diagonal ensemble":
\begin{equation}
    \lim_{T \rightarrow \infty}{\frac{1}{T}\int_0^T{dt\ \langle \wO(t) \rangle}} = \sumal{\alpha}{}{|c_\alpha|^2 O_{\alpha\alpha}}.     
\end{equation}
In a thermalizing system, we expect the long-time average to be equal to the expectation value of a local operator in a microcanonical ensemble around energy $\bar{E}$ of the initial state, requiring
\begin{equation}
    \sumal{E_\alpha \in \left[\bar{E} - \Delta, \bar{E} + \Delta\right]}{}{O_{\alpha\alpha}} = \sumal{\alpha}{}{|c_\alpha|^2 O_{\alpha\alpha}}.
\label{eq:MCtimeavg}
\end{equation}
Using the fact that magnitudes $|c_\alpha|^2$ are significant only in the energy window $[\bar{E} - \Delta, \bar{E} + \Delta]$, Eq.~(\ref{eq:MCtimeavg}) suggests that $O_{\alpha\alpha}$ on the RHS is only a function of the energy $\bar{E}$ rather than the eigenstate energy $E_\alpha$.  
These arguments, along with many other motivations~\cite{srednicki1994chaos}, led to a formal conjecture on the matrix elements of local operators in the energy eigenstates of a non-integrable model take the form~\cite{d2016quantum}
\begin{equation}
    \bra{E_m}\widehat{O}\ket{E_n} = \bar{O}\left(E\right)\delta_{m, n} + R_{m, n}\ \Omega(E)^{-1/2} f_O\left(E, \omega \right),
\label{eq:ethmatrixel}
\end{equation}
where $\widehat{O}$ is a local operator, $E = \left(E_m + E_n\right)/2$, $\omega = E_m - E_n$, $R_{m,n}$ is a pseudorandom variable such that the distribution of $\{R_{m,n}\}$ (over all values of $m$ and $n$) has zero mean and unit variance, $\bar{O}\left(E\right)$ is a smooth function of $E$ and represents the thermal expectation value of $\widehat{O}$ at energy $E$, $f_{O}\left(E, \omega\right)$ is a smooth function of $E$ and $\omega$ which do not scale with the system size~\cite{d2016quantum}, and $\Omega\left(E\right)$ is the density of states at energy $E$.
Note that the thermal value is typically determined in practice by computing the microcanonical average, i.e. averaging the eigenstate expectation values $\bra{E} \widehat{O} \ket{E}$ over a small energy window $\Delta$ that corresponds to the Thouless energy scale~\cite{beugeling2014finite}.
In Eq.~(\ref{eq:ethmatrixel}), for a system with Hilbert space dimension $\mathcal{D}$, we expect $\Omega(E) \sim 1/\mathcal{D}$ for states in the middle of the spectrum.
Hence, the standard deviation of expectation values of operators in the eigenstates in the middle of the spectrum within the Thouless energy window $\Delta$ is expected to scale as $\sim 1/\sqrt{\mathcal{D}}$, which forms a standard numerical diagnostic of ETH~\cite{beugeling2014finite} (although this scaling is debated~\cite{huang2022finite}). 
We refer to Eq.~(\ref{eq:ethmatrixel}) restricted to the cases $m = n$ and $m \neq n$ as diagonal ETH and off-diagonal ETH respectively~\cite{shiraishimorireply}. 
In this review,  we are primarily interested in the behavior of expectation values of local operators in eigenstates of the system, and hence in diagonal ETH.  
Note that for systems with additional symmetries such as particle number conservation, Eq.~(\ref{eq:ethmatrixel}) is expected to hold for eigenstates within a particular quantum number sector~\cite{rigol2008thermalization, d2016quantum, mondaini2018comment}. 
The question of which initial states thermalize under time-evolution leads to two notions of diagonal ETH: \textit{strong} ETH and \textit{weak} ETH. 
Strong ETH states that \textit{all} eigenstates obey diagonal ETH as stated in Eq.~(\ref{eq:ethmatrixel}), which implies that all initial states thermalize.
Evidence for the validity of strong ETH in typical non-integrable models has been found in Refs.~\cite{kim2014testing, garrison2018does}. 
On the other hand, weak ETH states that \textit{almost all} eigenstates obey diagonal ETH.  
In particular, this implies that there could be a small set (of fraction going to zero with increasing system size) of eigenstates violating diagonal ETH.
Such a scenario can in principle lead to the non-thermalization of a few special initial states amidst the thermalization of most initial states. 
We will discuss this scenario in more detail in Sec.~\ref{subsec:ergodicitybreaking}.
%

%
%
\subsection{Level Statistics}\label{subsec:levelstats}
Ergodicity in isolated quantum systems is typically considered synonymous with quantum chaos, a widely studied subject~\cite{stockmann2000quantumchaos}. 
A system is said to be quantum chaotic if its correlation functions under time-evolution by the Hamiltonian at late times resembles correlations under time-evolution by a Random Matrix with the same symmetries.
These considerations lead to defining features of quantum chaotic systems, such as the repulsion of nearest-neighbor eigenvalues~\cite{bgs1984chaos} and the linear ramp in the Spectral Form Factor (SFF) of such systems~\cite{brezin1997sff}.
Random Matrix Theory also provides remarkably accurate predictions of these quantities, and in particular for the statistics of nearest-neighboring energy differences $s_n = (E_{n+1} - E_n)/\bar{E}$, where $E_n$'s are the sorted energy levels and $\bar{E}$ is the mean energy level spacing in the vicinity of $E_n$~\cite{dyson1962brownian}.
It has been numerically verified for several non-integrable models that $s_n$ follows a Wigner-Dyson distribution~\cite{poilblanc1993poisson,rigol2008thermalization,nandkishore2015many} whereas $s_n$ in systems with several symmetries (e.g. integrable systems) follows a Poisson distribution~\cite{giraud2020probing}.
This distribution can also be directly detected using the mean level spacing ratio $\langle r \rangle$, which is the average of $r_n = \min(s_n, s_{n+1})/\max(s_n, s_{n+1})$~\cite{oganesyan2007localization, atas2013distribution}.
$\langle r \rangle \approx 0.53$ and $\langle r \rangle \approx 0.6$ for Wigner-Dyson ensembles with and without time-reversal symmetry, and $\langle r \rangle \approx 0.38$ for the Poisson distribution. 
Note that for non-integrable Hamiltonians with a few additional symmetries (e.g. particle number), signatures of ergodicity and its breaking are expected to appear in the distribution of energy levels within a symmetry sector~\cite{rigol2008thermalization, d2016quantum, mondaini2018comment}.
One common signature of the breakdown of ergodicity is hence the change in the distribution of level statistics after resolving known symmetries, as we will discuss in Sec.~\ref{subsec:ergodicitybreaking}.  
\subsection{Entanglement}\label{subsec:entanglement}
\begin{figure}
    \centering
    \includegraphics[scale=0.25]{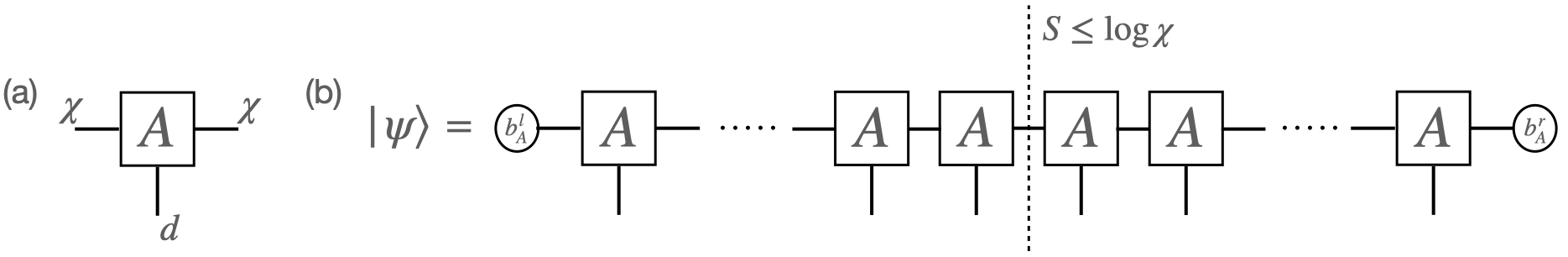}
    \caption{(a) $d \times \chi \times \chi$ tensor representing an MPS. $d$ is the physical dimension and $\chi$ is the bond-dimension (b) Wavefunction $\ket{\psi}$ represented in MPS form, $b^l_A$ and $b^r_A$ are $\chi$-dimensional boundary vectors. The entanglement entropy of an MPS state is bounded by $S \leq \log \chi$. }
    \label{fig:MPS}
\end{figure}
The concepts of entanglement and entropy (as one of its measure) are widely used in several contexts in physics~\cite{eisert2010colloquium, nishioka2009entanglement}, and are also crucial in the study of quantum dynamics as well as QMBS. 
The entanglement of a wavefunction $\ket{\psi}$ is defined via its Schmidt decomposition about a bipartition into regions $\ma$ and $\mb$, given by
\begin{equation}
    \ket{\psi} = \sumal{\alpha = 1}{\chi}{\lambda_\alpha \ket{\psi_\alpha}_{\ma} \ket{\psi_\alpha}_{\mb}}, 
\label{eq:bipartition}
\end{equation}
where $\{\ket{\psi_\alpha}_{\ma}\}$ and $\{\ket{\psi_\alpha}_{\mb}\}$ are orthonormal sets of wavefunctions on the subsystems $\ma$ and $\mb$ respectively, $\lambda_{\alpha}$'s are non-negative real numbers, and $\chi$ is known as the Schmidt rank of the wavefunction.  
For a normalized state $\ket{\psi}$, we always have $\sumal{\alpha = 1}{\chi}{\lambda_\alpha^2} = 1$. 
The (von Neumann) Entanglement Entropy (EE) $S$ of the state $\ket{\psi}$ over this bipartition is defined as 
\begin{equation}
    S \equiv -\sumal{\alpha = 1}{\chi}{\lambda_\alpha^2 \log \lambda_\alpha^2} = -\text{Tr}_{\mA}\left(\rho_{\ma} \log \rho_{\ma}\right) 
\label{eq:EEdefn}
\end{equation}
where $\rho_{\ma}$ is the reduced density matrix over subsystem $\ma$.  
Ground states of gapped quantum many-body systems are known to exhibit a so-called ``area-law" scaling of the EE, where $S$ scales proportionally to the area of the subsystem $\ma$, which, in one-dimension implies that $S$ is independent of the subsystem size.
On the other hand, the EE in ground states of critical gapless systems typically exhibit logarithmic violations of the area-law, i.e., $S$ scales with the area times the logarithm of the volume of the subsystem $\ma$~\cite{calabrese2009entanglement, eisert2010colloquium}.
For highly excited states of non-integrable models, ETH predicts a ``volume law" scaling of $S$ i.e. it scales proportionally to the volume of the subsystem $\mA$.
This is a direct consequence of the reduced density matrix discussed in Eq.~(\ref{eq:thermalization}).
In fact, for states in the middle of the spectrum, $\beta = 0$ in Eq.~(\ref{eq:thermalization}), and thus their EE is typically observed to be close to $S_{\textrm{th}}$, the mean EE of states in the Hilbert space~\cite{page1993}, also known as the Page entropy, which is close to the maximum possible entropy $S_{\textrm{max}}$. 
For a one-dimensional system with $L$ spin-$\frac{1}{2}$'s and $L_{\mA} = L/2$ spin-$\frac{1}{2}$'s in subsystem $\ma$, these values are known to be 
\begin{equation}
    S_{\textrm{th}} = \frac{L\log 2 - 1}{2},\;\;\; S_{\textrm{max}} = \frac{L \log 2}{2}.
\label{thermalent}
\end{equation}
Note that when $L_{\mA}/L$ is kept constant, $S_{\textrm{th}}$ and $S_{\textrm{max}}$ typically differ by an $L$-independent constant that only depends on the fraction $L_{\mA}/L$ and the properties of (e.g., the constraints on) the full Hilbert space~\cite{page1993, morampudi2020universal}; hence eigenstates in the middle of the spectrum are almost always maximally entangled. 
Entanglement also plays an important role in understanding the structure of QMBS eigenstates. 
For example, most of the QMBS eigenstates that we review have exact Matrix Product States (MPS) representations~\cite{schollwock2011density, verstraete2008matrix, orus2014practical, perezgarcia2007matrix}. 
An MPS wavefunction $\ket{\psi}$ (with open boundary conditions) can be written as
\begin{equation}
    \ket{\psi} = \sumal{\{m_1 m_2 \dots m_L\}}{}{[{b^l_A}^T A^{[m_1]} \dots A^{[m_L]} b^r_A] \ket{m_1 \dots m_L}},
\end{equation}
where $\ket{m_1 \dots m_L}$ denotes the many-body basis states where $\{m_j\}$ can take $d$ values, the dimension of the local physical Hilbert space. 
$\{A^{[m_j]}\}$'s are $\chi \times \chi$ matrices, where $\chi$ is referred to the bond-dimension of the MPS, and $b^l_A$ and $b^r_A$ are $\chi$-dimensional left and right boundary vectors that determine the boundary conditions for the wavefunction.  
Diagrammatically, $A$ can be visualized as a $d \times \chi \times \chi$ tensor as shown in Fig.~\ref{fig:MPS}a and the state $\ket{\psi}$ as contractions of these tensors shown in Fig.~\ref{fig:MPS}b.
Note that although any state can have multiple MPS representations, there is a canonical form of the MPS which has bond dimension $\chi$ that is the same as the number of non-zero Schmidt values of the state in Eq.~(\ref{eq:EEdefn}).
The EE for an MPS wavefunction then satisfies the bound
\begin{equation}
    S \leq \log \chi.
\label{eq:MPSEEbound}
\end{equation}
Hence the growth of the bond-dimension $\chi$ of an MPS representation of the state is sufficient to determine the scaling of EE with system-size. 
For example, it is well-known that area-law ground states of gapped systems in one dimension admit approximate/exact MPS representations of bond dimension $\chi$ that is system-size independent~\cite{hastings2007area}.
In Sec.~\ref{sec:towers}, we will apply these results to QMBS eigenstates and obtain the scaling of their EEs. 
\subsection{Ergodicity Breaking}\label{subsec:ergodicitybreaking}
Most local interacting Hamiltonians are believed to be non-integrable and fully ergodic, i.e. their eigenstates obey strong ETH.
Indeed, the spectrum of a generic local Hamiltonian exhibits level repulsion and Wigner-Dyson level statistics, signalling the presence of quantum chaos. 
Nevertheless, as discussed in Sec.~\ref{sec:intro}, systems with various degrees of ergodicity breaking have been found.
A complete breakdown of ergodicity, i.e. a breakdown of strong and weak ETH, is known in two types of systems: Integrable and Many-Body Localized (MBL). 
In addition, two types of partial breakdowns of ergodicity have been found, and the phenomena go by the names Quantum Many Body Scars (QMBS) and Hilbert space fragmentation. 
We now provide a brief overview these types of ergodicity breaking in isolated quantum systems, and a summary of their properties is provided in Table~\ref{tab:taxonomy}.
\subsubsection{Integrability}
Integrability occurs in the presence of an extensive number of conserved quantities, which leads to the complete solvability of the energy spectrum, at least in principle.
The simplest examples of integrable systems are non-interacting systems,  where the many-body spectrum is completely determined by the single-particle spectrum. 
Interacting examples of integrable models include ones with commuting projector Hamiltonians, such as the toric code~\cite{kitaev2003fault, kitaev2010topological}, being a celebrated example, and Bethe ansatz solvable models such as the one-dimensional XXZ and the one-dimensional Hubbard models~\cite{faddeev1996algebraic}.
Quantum integrability is not stable under generic perturbations, and moreover no analogues of the Kolmogorov-Arnold-Moser (KAM) theorem for classical integrability, that establishes some degree of stability under perturbations, have been rigorously established for quantum systems. 
Hence it is believed that a high degree of fine-tuning is required in the space of all local Hamiltonians in order to see signatures of quantum integrability.
\subsubsection{Many-Body Localization}
On the other hand,  MBL, the generalization of Anderson localization to interacting systems, is believed to occur more generically in the presence of strong disorder or quasiperiodicity~\cite{nandkishore2015many, abanin2018review}, although its stability in the thermodynamic limit has been a subject of active debate~\cite{deroeck2017stability, suntajs2020quantum, sels2020dynamical, crowley2020constructive}.  
In both cases, the existence of an extensive number of conserved quantities can be constructed, which leads to the absence of level repulsion that is reflected in the Poisson level statistics shown by these Hamiltonians.  
From the point of view of entanglement, MBL systems possess eigenstates with area-law entanglement that are easy to identify~\cite{huse2014phenomenology}.
On the other hand, quantum integrable systems mostly possess volume-law entangled eigenstates with a few exceptions~\cite{alba2009excited, vidmar2017entanglement}, and their eigenstates are hence harder to distinguish from thermal eigenstates~\cite{alba2015eigenstate, khemani2014eigenstate}.
\subsubsection{Quantum Many-Body Scars (QMBS)}
A distinct type of ergodicity breaking, termed as \textit{weak ergodicity breaking} in Ref.~\cite{turner2017quantum} can occur in systems that violate strong ETH but still obey weak ETH. 
Such systems exhibit a few highly excited eigenstates that violate diagonal ETH, i.e. they possess atypical features compared to most other eigenstates at the same energy density. 
These ETH-violating eigenstates in the middle of the spectrum are referred to as \textit{Quantum Many-Body Scars} (QMBS). 
The term originates from the analogy to quantum scars in single-particle systems such as a Bunimovich stadium~\cite{heller1984scars} or quantum maps~\cite{backer2003numerical}, where a small set of single-particle eigenstates with anomalous wavefunctions distributed on rare classical periodic orbits co-exist with generic eigenstates with wavefunctions distributed uniformly.
Such systems are said to violate the Quantum Unique Ergodicity (QUE) conjecture, which is, roughly speaking, the analogue of strong ETH for single-particle systems.
Typically, the number of QMBS grows exponentially slower than the Hilbert space dimension (either polynomially in system size or exponentially with a smaller base), and they constitute a measure-zero set in the thermodynamic limit.
Since most of the spectrum exhibits level repulsion, systems with QMBS show level repulsion on average, as well as many other standard signatures of quantum chaos. 
We might expect that signatures of QMBS buried in the middle of the spectrum would be hard to experimentally access.
Nevertheless, systems in which QMBS appear as equally spaced towers in the spectrum are of particular interest since equal spacings result in perfect revivals from particular initial states, a phenomenon that has been observed in Rydberg atom experiments~\cite{bernien2017probing, bluvstein2020controlling}.  
We discuss systems with equally spaced towers of QMBS in Sec.~\ref{sec:towers} and associated unified formalisms in Sec.~\ref{sec:unifiedformalism}, and systems with isolated QMBS in Sec.~\ref{sec:isolated}.
%
%
\subsubsection{Hilbert Space Fragmentation}
Another type of ergodicity breaking of a different origin can occur in constrained systems, where the Hilbert space splits into exponentially many dynamically disconnected parts, such that large parts of it are inaccessible to particular initial states.
The term \textit{Hilbert space fragmentation} was coined in Ref.~\cite{sala2019ergodicity} to refer to such systems.
Fragmentation was divided into two main categories: \textit{weak} and \textit{strong}, depending on whether the fraction of states violating the conventional form of ETH are a set of measure zero or not in the thermodynamic limit.
Weakly fragmented systems are also sometimes regarded as examples of QMBS~\cite{sala2019ergodicity, khemani2019local}, since they obey weak ETH since the ETH-violating states form a set of measure-zero.
However, strongly fragmented systems also violate conventional forms of weak ETH, and should be regarded as a distinct form of ergodicity breaking.
Fragmented systems possess eigenstates that can show any scaling of EE from area-law to volume-law, depending on the size of the dynamically disconnected part of the Hilbert space they belong to. 
Moreover, as we discuss in more detail in Sec.~\ref{sec:fragmentation}, while weakly fragmented systems exhibit Wigner-Dyson level statistics, strongly fragmented systems typically consist of a large number of degeneracies in the spectrum, and can exhibit unconventional level statistics~\cite{giraud2020probing, herviou2021manybody}.
%
%
\begin{table}[]
    \centering
    \begin{tabular}{|c|c|c|c|}
    \hline
          & {\bf Strong/Weak ETH} &  {\bf Entanglement} & {\bf Level Statistics} \\
         \hline
        {\bf Ergodic}  & Yes/Yes  & Volume & Wigner-Dyson\\
        \hline
        {\bf Integrable}  & No/No & Volume/Sub-Volume & Poisson\\
        \hline
        {\bf MBL}  & No/No & Area & Poisson\\
        \hline
        {\bf Quantum Scarred}  & \multirow{2}{*}{No/Yes} & \multirow{2}{*}{Volume/Sub-Volume} & \multirow{2}{*}{Wigner-Dyson}\\
        {\bf Weakly Fragmented} &  &  &  \\
        \hline
        {\bf Strongly Fragmented}  & No/No &  Volume/Sub-Volume & Poisson\\
        \hline
    \end{tabular}
    \caption{Taxonomy of Ergodicity and various types of its breaking in Isolated Quantum Systems. They can be distinguished based on whether they satisfy strong/weak ETH, the entanglement entropy scaling of typical eigenstates in the middle of the spectrum, and their energy level statistics.}
    \label{tab:taxonomy}
\end{table}
\section{Towers of QMBS}\label{sec:towers}
Since the ETH-violating eigenstates in quantum scarred systems constitute a measure-zero set in the thermodynamic limit, it is natural to wonder whether they would influence the dynamics of any experimentally accessible initial states.
As we discuss now, in cases where the spectrum includes an extensive number of non-thermal eigenstates in an equally spaced tower with energies $\{E_0, E_0 + \mE,  E_0 + 2\mE , \cdots,  E_0 + (N-1) \mE\}$,  novel dynamical phenomena are possible.
For example, the presence of such a tower in the spectrum leads to perfect revivals in the systems under dynamics from particular initial states. 
Revivals can be probed by computing the fidelity $\mF(t)$ (also known as Loschmidt echo) of an initial state $\ket{\psi(0)} = \sumal{n}{}{c_n \ket{E_n}}$,  defined as 
\begin{equation}
\mF(t) = |\braket{\psi(0)}{\psi(t)}|^2 = |\sumal{n}{}{|c_n|^2 e^{-i E_n t}}|^2 = \sumal{m, n}{}{|c_n c_m|^2 e^{i (E_m - E_n) t}}.
\label{eq:fidelity}
\end{equation}
For any initial state $\ket{\psi(0)}$ that lies completely within the subspace spanned by the tower of eigenstates,  all of the energy differences $\{E_m - E_n\}$ that appear in the sum of Eq.~(\ref{eq:fidelity}) are integer multiples of the spacing $\mE$, and hence the system exhibits perfect revivals with time-period of at most $T = \frac{2\pi}{\mE}$ (i.e., $\mF(t + T) = \mF(t)$).  
While this is obvious if $\ket{\psi(0)}$ is explicitly chosen to be a superposition of a few eigenstates $\{\ket{E_n}\}$, it is not always clear that a product state or any physically relevant state can be constructed that way. 
Nevertheless, as we will discuss in Sec.~\ref{subsec:revivals}, for models exhibiting QMBS, it is possible to construct states $\ket{\psi(0)}$ with area law entanglement that have overlap with an extensive number of eigenstates, and lie completely within the QMBS subspace.
Conversely, it has been shown on general grounds~\cite{alhambra2020revivals} that the existence of perfect revivals from a low-entanglement and short-range correlated state implies the presence of equally-spaced (or commensurately-spaced) eigenstates in the middle of the spectrum having low entanglement entropy, i.e., towers of QMBS.
The first exact, i.e., analytical, example of such a tower of eigenstates was found in the integer spin AKLT models~\cite{moudgalya2018a, moudgalya2018b}, well-known in the context of ground state and low-energy physics~\cite{aklt1987rigorous, arovas1988extended, arovas1989two}.
Subsequently, numerous works found similar towers in simpler models such as the spin-1 XY model~\cite{iadecola2018exact}, and connections were established to the phenomenon of $\eta$-pairing known in the context of Hubbard models~\cite{yang1989eta}. 
In all these examples, the states in the tower are composed of multiple quasiparticles of a given energy and momentum dispersing on top of a fixed low-entanglement eigenstate such as the ground state.
Several examples bear a direct resemblance to $\eta$-pairing in Hubbard models, and we discuss them in Sec.~\ref{subsec:SGAtowers}.   
We survey other examples of towers of states, some that appear to have a more complicated origin such as the AKLT model, as well as other miscellaneous examples in Sec.~\ref{subsec:othertowers}.
\subsection{Simple examples: Spectrum Generating Algebras}\label{subsec:SGAtowers}
In the following, we discuss some simple examples of models that exhibit exact QMBS, based on structures referred to as Spectrum Generating Algebras (SGA).
These examples form the foundation for systematic approaches to construct models with towers of exact QMBS.
They capture several examples of towers that have been discussed in the literature, which we will present in Sec.~\ref{sec:unifiedformalism}. 
\subsubsection{Hubbard model}
For pedagogical purposes, we explicitly illustrate the towers of exact eigenstates in the celebrated Hubbard model, known as $\eta$-pairing eigenstates~\cite{yang1989eta, yang1990so}.
The Hamiltonian for the Hubbard model on a hypercubic lattice in $d$-dimensions is given by
\begin{equation}
    \hhub = \sumal{\sigma \in \{\uparrow, \downarrow\}}{}{\left[-\sumal{\langle \vr, \vr'\rangle}{}{t_{\vr, \vr'} \left(\cd_{\vr, \sigma} c_{\vr', \sigma} + h.c.\right)} -\mu\sumal{\vr}{}{\hn_{\vr, \sigma}}\right]} + U \sumal{\vr}{}{\hn_{\vr, \uparrow} \hn_{\vr, \downarrow}},
\label{eq:FermiHubbard}
\end{equation}
where $\{\vr\}$ labels the sites of a lattice, $\langle \vr, \vr'\rangle$ denotes neighboring sites, and $t_{\vr, \vr'}$ denotes the corresponding hopping strength, which is typically chosen to be site-independent.
$\cd_{\vr, \sigma}$ and $c_{\vr,\sigma}$ denote the fermion creation and annihilation operators on site $\vr$, and the on-site fermion number operator is defined as $\hn_{\vr,\sigma} \equiv \cd_{\vr, \sigma} c_{\vr, \sigma}$.
The Hubbard model on any bipartite lattice has spin and pseudospin symmetries, which are examples of conventional and dynamical $SU(2)$ symmetries respectively, as we discuss below.
The spin $SU(2)$ symmetry is composed of the operators $\{S^+, S^-, S^z\}$ and the corresponding quadratic Casimir operator $\vec{S}^2$, which are defined as
\begin{gather}
    S^+ = \sumal{\vr}{}{\cd_{\vr, \uparrow} c_{\vr, \downarrow}},\;\;S^- = (S^+)^\dagger, \;\; S^z = \frac{1}{2}\sumal{\vr}{}{(\hn_{\vr, \uparrow} - \hn_{\vr, \downarrow})}\nn \\
    \vec{S}^2 = \frac{1}{2}(S^+ S^- + S^- S^+) + (S^z)^2.
\label{eq:spinsu2}
\end{gather}
Similarly, the pseudospin $SU(2)$ symmetry is composed of the operators $\{\ed, \eta, \eta^z\}$ and the corresponding quadratic Casimir $\vec{\eta}^2$, which are defined on a bipartite hypercubic lattice with $L$ sites (and even number sites in directions with periodic boundary conditions) as
\begin{gather}
    \ed = \sumal{\vr}{}{e^{i\bp\cdot\vr} \cd_{\vr, \uparrow} \cd_{\vr,\downarrow}},\;\;\eta = (\ed)^\dagger,\;\;\eta^z = \frac{1}{2}\left(\sumal{\vr, \sigma}{}{\hn_{\vr, \sigma}} - L\right)  \nn \\
    \vec{\eta}^2 \equiv \frac{1}{2}(\ed\eta + \eta \ed) + (\eta^z)^2.
\label{eq:etaops}
\end{gather} 
The spin and pseudospin are examples of $SU(2)$ symmetries since they obey the usual $\mathfrak{su}(2)$ commutation relations
\begin{equation}
    [S^z, S^+] = S^+,\;\;[S^z, S^-] = -S^-,\;\;[\eta^z, \ed] = \ed,\;\;[\eta^z, \eta] = -\eta.
\label{eq:su2comm}
\end{equation}
Further, they are symmetries of the Hubbard model of Eq.~(\ref{eq:FermiHubbard}) on a bipartite lattice since they satisfy
\begin{equation}
    [\hhub, \eta^z] = 0,\;\;[\hhub, \vec{\eta}^2] = 0,\;\;[\hhub, S^z] = 0,\;\;[\hhub, \vec{S}^2] = 0.
\label{eq:su2syms}
\end{equation}
As a consequence of Eq.~(\ref{eq:su2syms}), the eigenstates of the Hubbard model can be labelled by quantum numbers corresponding to two $SU(2)$ symmetries -- $(\vec{\eta}^2, \eta^z)$ and $(\vec{S}^2, S^z)$, although these quantum numbers are not completely independent of each other~\cite{essler2005one}. 
A crucial difference between the two $SU(2)$ symmetries lies in the commutation relation of $\hhub$ with $\ed$ and $S^+$, which read
\begin{equation}
    [\hhub, S^+] = 0,\;\;\;[\hhub, \ed] = (U - 2\mu)\ed. 
\label{eq:HubbardSGA}
\end{equation}
While the spin-$SU(2)$ is an example of a conventional $SU(2)$ symmetry, the latter is referred to as a \textit{Spectrum Generating Algebra} (SGA) or a \textit{Dynamical Symmetry}~\cite{barut1965dynamical, dothan1965series, buca2019nonstationary, medenjak2020isolated}, when for a Hamiltonian $H$ an operator $\ed$ satisfies
\begin{equation}
    [H, \ed] = \mE \ed.
\label{eq:SGA}
\end{equation}
The conventional $SU(2)$ symmetry is a special case of Eq.~(\ref{eq:SGA}) where $\mE = 0$. 
While a conventional $SU(2)$ symmetry results in the existence of degenerate multiplets of states in the spectrum (which are related by the action of raising and lowering operators $\ed$ and $\eta$), an SGA with $\mE \neq 0$ leads to the existence of a tower of equally spaced energy eigenstates, i.e. if $\ket{\psi_0}$ is an eigenstate of $H$ with energy $E_0$, $\ed \ket{\psi_0}$ is also an eigenstate with energy $E_0 + \mE$.
Choosing $\ket{\psi_0}$ to be an eigenstate of the Casimir operator $\vec{\eta}^2$ and $\eta^z$ with eigenvalues $J(J+1)$ and $-J$ respectively for some $J$, we obtain a multiplet of $(2J + 1)$ eigenstates 
\begin{equation}
    \{\ket{\psi_0}, \ed \ket{\psi_0}, \cdots, (\ed)^{2J} \ket{\psi_0}\} 
\label{eq:psi0tower}
\end{equation}
with equally spaced energies given by
\begin{equation}
    \{E_0, E_0 + \mE, E_0 + 2 \mE, \cdots, E_0 + 2 J \mE\}.
\end{equation}
Provided the state $\ket{\psi_0}$ is a solvable eigenstate, Eq.~(\ref{eq:psi0tower}) denotes an exact tower of eigenstates. 
A special set of solvable eigenstates of the Hubbard model are spin-polarized states that consist of only one type of spin $\uparrow$ or $\downarrow$.
The interaction term in Eq.~(\ref{eq:FermiHubbard}) acts trivially on these states, which enables the construction of subspaces in which the action of the Hubbard model reduces to a quadratic Hamiltonian that can be solved exactly.
The simplest example of a solvable state is the vacuum state $\ket{\Omega}$ with no particles, and it can be used to construct a simple exact tower of states of the form Eq.~(\ref{eq:psi0tower})~\cite{vafek2017entanglement}.
For example, such a tower in one dimension has the following form
\begin{gather}
    \ket{\Omega} = \ket{0\ 0\ \cdots\ 0\ 0},\;\;\ed \ket{\Omega} =\sumal{j}{}{(-1)^j\ \overset{j}{\ket{0\ \cdots\ 0\ \updownarrow\ 0\ \cdots\ 0}}},\nn \\
    \;\;(\ed)^2\ket{\Omega} = \sumal{j,k}{}{(-1)^{j+k}\overset{j\hspace{13.5mm}k}{\ket{0 \cdots 0 \updownarrow 0 \cdots 0 \updownarrow 0 \cdots 0}}},\;\;\cdots\nn \\
    \cdots,\;\;(\ed)^L\ket{\Omega} = \ket{\updownarrow\ \updownarrow\ \cdots\ \updownarrow\ \updownarrow}, 
\label{eq:HubbardQP}
\end{gather}
where $0$ denotes an empty site, $\uparrow$ and $\downarrow$ denote sites with one of the two types of spins, and $\updownarrow$ denotes a doubly occupied site, which we refer to as a doublon.
As evident from Eq.~(\ref{eq:HubbardQP}), the state $(\eta^\dagger)^n\ket{\Omega}$ consists of $n$ momentum $\pi$ doublon ``quasiparticles" dispersing around the system, so that the full state has momentum $n \pi$. 
As we will discuss later in Secs.~\ref{subsec:othertowers} and \ref{sec:unifiedformalism}, the quasiparticle nature of eigenstates is a general feature of towers of QMBS. 
Several of the analytically tractable towers of states in the Hubbard model, including the ones of Eq.~(\ref{eq:HubbardQP}) do not exhibit a volume-law scaling of EE~\cite{vafek2017entanglement}, owing to their quasiparticle nature. 
While some of these towers are also in the middle of the full many-body energy spectrum, to really qualify as examples of QMBS, they should be in the middle of the spectrum \textit{after} resolving symmetries of the system~\cite{d2016quantum, mondaini2018comment}.
However, for the simplest tractable towers such as Eq.~(\ref{eq:HubbardQP}), it turns out that they are the only states within their quantum number sector after resolving the spin and pseudospin $SU(2)$ symmetries of Eq.~(\ref{eq:su2syms})~\cite{vafek2017entanglement}, hence they are not considered as examples of QMBS in the Hubbard model.\footnote{There are other analytically tractable towers in $\hhub$ obtained by the repeated action of $\ed$ on certain spin-polarized eigenstate of $\hhub$, which are not the only ones in their quantum number sector, are in the middle of the spectrum, and have a sub-volume law scaling of EE~\cite{vafek2017entanglement}. These should be considered examples of QMBS in the Hubbard model.}
Nevertheless, Refs.~\cite{moudgalya2020eta, mark2020eta} showed that local terms can be added to the Hubbard model that break either one of the two $SU(2)$ symmetries and translation symmetry while preserving some of the analytically tractable towers as eigenstates.
Such models are said to exhibit a Restricted Spectrum Generating Algebra (RSGA)~\cite{moudgalya2020eta}, and we will discuss its precise statement in Sec.~\ref{sec:unifiedformalism}.
The remaining towers of states in such models are generically in the middle of the spectrum after resolving all the conventional symmetries of the model, and hence are examples of towers of QMBS. 
For example, the states of the vacuum tower $\{(\eta^\dagger)^n \ket{\Omega}\}$ are exact eigenstates for the Hubbard model with additional electrostatic terms, i.e.,
\begin{equation}
    H_{\scar} = \hhub + \sumal{\dlangle \vr, \vr'\drangle}{}{V_{\vr,\vr'} S^z_{\vr} S^z_{\vr'}},
\label{eq:Int}
\end{equation}
where $V_{\vr, \vr'}$'s are arbitrary real coefficients.
This Hamiltonian $H_{\scar}$ breaks the $SU(2)$ symmetries of $\hhub$, and hence the states of the tower of Eq.~(\ref{eq:HubbardQP}) are examples of QMBS eigenstates of $H_{\scar}$.
An exhaustive search of such nearest-neighbor terms that preserve $\{(\eta^\dagger)^n \ket{\Omega}\}$ as exact eigenstates was performed in Ref.~\cite{mark2020eta}, which includes the Hamiltonian of Eq.~(\ref{eq:Int}).
Further, Ref.~\cite{moudgalya2020eta} identified some such terms and provided sufficient conditions for the preservation of such towers originating from an SGA, a formalism we will briefly discuss in Sec.~\ref{sec:unifiedformalism}.

\subsubsection{Ferromagnetic Towers}
Notice that the SGA or dynamical symmetry is similar to systems where an $SU(2)$ symmetry is ``broken" by a constant magnetic field.
Indeed, starting with an $SU(2)$-symmetric Hamiltonian $H_0$, adding magnetic field $B$ results in an SGA for the usual spin $SU(2)$ symmetry
\begin{equation}
    H_B = H_0 + B S^z,\;\;[H_B, S^+] = B S^+,\;\;[H_B, \vec{S}^2] = 0,\;\;[H_B, S^z] = 0. 
\label{eq:magneticSGA}
\end{equation}
For example, $H_0$ can be the Hubbard model of Eq.~(\ref{eq:FermiHubbard}),  or the one-dimensional spin-1/2 Heisenberg model given by
\begin{equation}
    H_0 = \sumal{j}{}{J_j \vec{S}_j\cdot\vec{S}_{j+1}}.
\label{eq:heisenberg}
\end{equation}
These Hamiltonians admit an exactly solvable ``ferromagnetic" states of the form $\ket{\psi_0} = \ket{\downarrow \cdots \downarrow}$, say with energy $E_0$.
The SGA condition of Eq.~(\ref{eq:magneticSGA}) ensures the existence of a ``ferromagnetic tower" of eigenstates $\{(S^+)^n \ket{\psi_0}\}$, which are exact eigenstates of the Hamiltonian $H_B$ with energies $\{E_0 + B n\}$ for $0 \leq n \leq L + 1$. 
While the states in this ferromagnetic tower of $H_B$ are not considered scars of the Hamiltonian, local perturbations that break the SGA of Eq.~(\ref{eq:magneticSGA}) (for example, ones that that do not commute with $\vec{S}^2$) can be added that preserved the eigenstates of the ferromagnetic tower, leading to the features we discussed for the vacuum tower of the Hubbard model of Eq.~(\ref{eq:FermiHubbard}). 
An exhaustive search for such types of terms was performed in Ref.~\cite{mark2020eta}.
This led to the discovery of physically relevant models with the ferromagnetic tower as QMBS, including some terms $H_{\textrm{DMI}}$ with a Dzyaloshinskii-Moriya Interaction (DMI)~\cite{dzyaloshinsky1958thermodynamic, moriya1960anisotropic, dooley2021robust}, e.g., 
\begin{equation}
    H_{\textrm{DMI}} = \sumal{j }{}{\hat{z}\cdot\left(\vec{S}_j \times \vec{S}_{j+1}\right)} = \sumal{j }{}{\left(S^x_j S^y_{j+1} - S^y_j S^x_{j+1}\right)}.
\label{eq:DMIexample}
\end{equation}
\subsubsection{Spin-1 XY Model}
\begin{figure}
    \centering
    \includegraphics[scale=1]{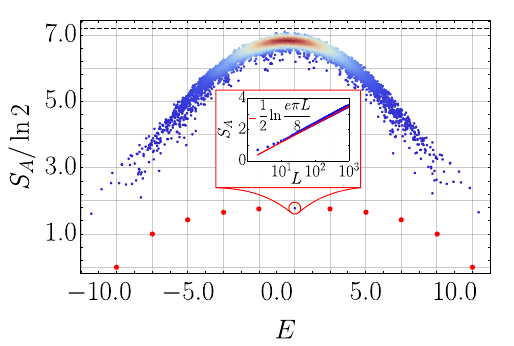}
    \caption{(Color online) Bipartite Entanglement Entropy (EE) $S_A$ as a function of eigenstate energy $E$ of the one-dimensional spin-1 XY model of Eq.~(\ref{eq:spin1XY})(figure reproduced from Ref.~\cite{schecter2019weak}).  Data  shown for a system of size $L = 10$ with the parameters $(h, D, J) = (1, 0.1, 0.1)$ with OBC within the sector zero total spin.  
    The red dots depict the EE for the QMBS eigenstates, which lie in other quantum number sectors (apart from the circled state at $E = 1$), but nevertheless are clear outliers among other eigenstates in the spectrum.
     Inset: Logarithmic scaling of the EE of the circled QMBS eigenstate as a function of the system size.
    }
    \label{fig:scarproperties_ent}
\end{figure}

Similar towers of QMBS are found in the spin-1 XY model~\cite{schecter2019weak}, given by the Hamiltonian
\begin{equation}
    H_{\textrm{XY}} = J \sumal{\langle \vr, \vr' \rangle}{}{\left(S^x_{\vr} S^x_{\vr'} + S^y_{\vr} S^y_{\vr'}\right)} + h \sumal{\vr}{}{S^z_{\vr}} + D \sumal{\vr}{}{(S^z_\vr)^2},
\label{eq:spin1XY}
\end{equation}
where $\langle \vr, \vr' \rangle$ denote nearest-neighboring sites on a lattice, $\{S^\alpha_{\vr}\}$, $\alpha \in \{x, y, z\}$ denote the spin-1 operators on site $\vr$.
On a general lattice, $H_{\textrm{XY}}$ only has a conventional $U(1)$ symmetry generated by the total spin operator $\sum_{\vr}{S^z_{\vr}}$, but it possesses simple spin-polarized such as $\ket{\Omega} = \ket{- \cdots -}$ and $\ket{\bar{\Omega}} = \ket{+ \cdots +}$, where $-$ and $+$ denote on-site spin configurations with $S^z = -1$ and $S^z = +1$ respectively.
Ref.~\cite{schecter2019weak} showed the existence of a tower of QMBS that connects the states $\ket{\Omega}$ and $\ket{\bar{\Omega}}$, of the form $\{(Q^\dagger)^n \ket{\Omega}\}$ with energies $\{2 h n\}$, where $Q^\dagger \equiv \sumal{\vr}{}{e^{i \bp\cdot \vr} (S^+_{\vr})^2}$ is the quasiparticle creation operator, similar to $\ed$ of Eq.~(\ref{eq:etaops}). 
Indeed, in one dimension, the tower of states consists of multiple quasiparticles with momentum $\pi$, similar to Eq.~(\ref{eq:HubbardQP})
\begin{gather}
    \ket{\Omega} = \ket{-\ -\ \cdots\ -\ -},\;\;Q^\dagger \ket{\Omega} =\sumal{j}{}{(-1)^j\ \overset{j}{\ket{-\ \cdots\ -\ +\ -\ \cdots\ -}}},\nn \\
    \;\;(Q^\dagger)^2\ket{\Omega} = \sumal{j,k}{}{(-1)^{j+k}\overset{j\hspace{15.5mm}k}{\ket{- \cdots - + - \cdots - + - \cdots -}}},\;\;\cdots\nn \\
    \cdots,\;\;(Q^\dagger)^L\ket{\Omega} = \ket{\bar{\Omega}} = \ket{+\ +\ \cdots\ +\ +}, 
\label{eq:spin1XYQP}
\end{gather}
As discussed in Ref.~\cite{schecter2019weak}, the states of this tower are generically in the middle of the spectrum of the $H_{\textrm{XY}}$ after resolving all the conventional symmetries of the model,  and they have a sub-volume scaling of the entanglement entropy, (as we also discuss in Sec.~\ref{subsec:QPentanglement}), hence they form examples of QMBS.
These features are also evident in Fig.~\ref{fig:scarproperties_ent},  where we show the entanglement entropy of the eigenstates of the spin-1 XY model within a quantum number sector along with those for the QMBS eigenstates.
These QMBS are similar to those in $H_{\scar}$ of Eq.~(\ref{eq:HInt}), and the $D = 0$ limit of the Hamiltonian of Eq.~(\ref{eq:spin1XY}) admits an SGA with the operator $Q^\dagger$, similar to the Hubbard model (see Eq.~(\ref{eq:HubbardSGA})).
In fact, Ref.~\cite{mark2020eta} established an exact correspondence between the tower of QMBS of Eq.~(\ref{eq:HubbardQP}) in the Hubbard and related models and tower of QMBS in the spin-1 XY model of Eq.~(\ref{eq:spin1XYQP}).
In particular, the lowest and highest eigenstates of the towers are respectively identified (i.e., $\ket{- \cdots -}$ and $\ket{+ \cdots +}$ correspond to $\ket{0 \cdots 0}$ and $\ket{\updownarrow \cdots \updownarrow}$), and the raising operator $Q^\dagger$ in the spin-1 XY model was identified with the $\eta^\dagger$ in the Hubbard model.
\subsection{Survey of other towers in the literature}\label{subsec:othertowers}
\subsubsection{AKLT Model}
The examples of towers of QMBS discussed in Sec.~\ref{subsec:SGAtowers} share the property that the raising and lowering operators for the towers of states were Hermitian conjugates of each other, a property that is a direct consequence of the underlying SGA of Eq.~(\ref{eq:SGA}).
However, not all towers of QMBS in the literature have this property.
In fact the first example of a tower of QMBS, found in the spin-1 AKLT model in Ref.~\cite{moudgalya2018a}, violates this condition. 
The AKLT model consists of $L$ spin-1's, and its Hamiltonian reads
\begin{equation}
    \haklt = \sum_{j}{P^{(2)}_{j, j+1}} = \sum_{j}{\left(\frac{1}{3} + \frac{1}{2}\left(\vec{S}_j\cdot\vec{S}_{j+1}\right) + \frac{1}{6}(\vec{S}_j\cdot\vec{S}_{j+1})^2\right)},
\label{eq:AKLTHamiltonian}
\end{equation}
where the nearest-neighbor terms $P^{(2)}_{j, j+1}$ are projectors of two neighboring spin-1's on sites $j$ and $j+1$ into total angular momentum spin-$2$ state, the sum over $j$ runs from $1$ to $L$ or $L-1$ for periodic boundary conditions (PBC) or open boundary conditions (OBC) respectively, and the position subscripts are modulo $L$ for PBC.
The ground state $\ket{G}$ of $\haklt$ of Eq.~(\ref{eq:AKLTHamiltonian}), also sometimes referred to as the AKLT state, is a frustration-free ground state, i.e. $P^{(2)}_{j, j+1}\ket{G} = 0$ for any $j$, and it can be elegantly expressed in terms of Schwinger bosons and dimers~\cite{aklt1987rigorous}.
Exact expressions for several excited states in $\haklt$ were first constructed in the language of dimers in Ref.~\cite{moudgalya2018a}, following the construction of two exact low-energy eigenstates in Ref.~\cite{arovas1989two}. 
While many such excited states are energetically close to the edge of the spectrum, an equally-spaced tower of exact eigenstates with energies in the bulk of the spectrum was obtained for even system sizes with PBC and for all system sizes with OBC.
The states of this tower are composed of multiple non-interacting quasiparticles dispersing with momentum $k = \pi$ (for PBC) on top of the ground state $\ket{G}$, similar to the tower of Eq.~(\ref{eq:HubbardQP}) in the Hubbard model $\hhub$.
The quasiparticle creation operator for this tower in $\haklt$ reads $Q^\dagger = \sum_j{(-1)^j (S^+_j)^2}$, and the states $\{(Q^\dagger)^n \ket{G}\}$ are eigenstates of $\haklt$ with energies $\{E = 2n\}$, total spin $\{s = 2n\}$, and its $z$-projection $\{S_z = 2n\}$.
Hence they form an extensive tower of states starting from the ground state $\ket{G}$ with energy $E = 0$ to the highest excited ``ferromagnetic" state $\ket{F}$ with energy $E = L$. 
These eigenstates reside in the bulk of the spectrum \textit{after} resolving all known symmetries of $\haklt$, which include translation (for PBC), inversion, and $SU(2)$~\cite{moudgalya2018a}.
Moreover, these states obey a sub-volume-law scaling of their entanglement entropy, owing to their quasiparticle structure~\cite{moudgalya2018b}, as we will show in Sec.~\ref{subsec:QPentanglement}. 
Note that as a consequence of the $SU(2)$ symmetry of $\haklt$, there is a multiplet of $(4n+1)$ eigenstates associated with each ``highest-weight" state $(Q^\dagger)^n\ket{G}$.
The ``lowest-weight" states of the multiplet read $Q^n\ket{G}$, where $Q = \sum_j{(-1)^j (S^-_j)^2}$, and these are eigenstates of $\haklt$ with energies $\{E = 2n\}$, total spin $\{s = 2n\}$, and its $z$-projection $\{S_z = -2n\}$. 
Hence, unlike the towers discussed in Sec.~\ref{subsec:SGAtowers}, the $Q$ operator is not a lowering operator for the tower created by the action of the $Q^\dagger$ operator, but they create the lowest and highest states of a multiplet. 
The lowering operator for the tower created by $Q^\dagger$ is believed to be a distinct (possibly non-local) operator $Q'$, as shown in Fig.~\ref{fig:towerpyramid}.
\subsubsection{General Structure of QMBS Towers}
We now comment on some general features of QMBS towers that appear in the literature.
As evident from the above examples, towers of QMBS are comprised of ``non-interacting" quasiparticles on top a product state (or more generally, on an MPS eigenstate with finite bond-dimension).
Heuristically, the interaction between quasiparticles can be forbidden in two distinct ways, which is one distinguishing feature between the QMBS in the AKLT model and those discussed in Sec.~\ref{subsec:SGAtowers}.
First, interactions can be forbidden when there are no terms in the Hamiltonian that energetically penalize configurations of nearest-neighbor quasiparticles, which occurs in the examples discussed in Sec.~\ref{subsec:SGAtowers}.
For example, in the $\eta$-pairing tower of Eq.~(\ref{eq:HubbardQP}), the energy of the configuration $\ket{\updownarrow\ \updownarrow}$ with quasiparticles beside each other is not affected by any of the terms in the Hamiltonian $H_{\scar}$. 
On the other hand, the quasiparticles can have ``emergent kinetic constraint"~\cite{iadecola2020quantum}, where configurations with quasiparticles beside each other do not appear in the wavefunctions of states of the tower.
Such a scenario occurs, for example, in the quasiparticle in the QMBS of the AKLT model, where the quasiparticle creation operator satisfies $(S^+_j)^2 (S^+_{j+1})^2\ket{G} = 0$, implicitly disallowing any configuration where quasiparticles are beside each other.
Towers of QMBS with this phenomenology were also found in the spin-$S$ $SO(3)$-symmetric AKLT models~\cite{moudgalya2018a, moudgalya2018b}, spin-$S$ $SO(2S + 1)$-symmetric AKLT models in Ref.~\cite{odea2020from}, and the towers of states in certain Domain-Wall-Conserving (DWC) models studied in Ref.~\cite{iadecola2020quantum}, which share several features with the QMBS in the AKLT models. 
These microscopic properties of quasiparticles can be used to systematically construct Hamiltonians with QMBS.
For examples, such properties were utilized in Ref.~\cite{moudgalya2020large} to systematically construct scarred models based on parent Hamiltonians of Matrix Product States (MPS)~\cite{perezgarcia2007matrix}.
In particular, this led to the discovery of a 6-parameter family of Hamiltonians with $\{(Q^\dagger)^n\ket{G}\}$ as QMBS, of which $\haklt$ was a special case (The same family was independently discovered in Ref.~\cite{mark2020unified} using a different approach, which we will discuss in Sec.~\ref{subsec:SGAbased}).  
Similar approaches can be used to systematically construct families of Hamiltonians with the same tower of QMBS as the spin-1 XY model (see Eq.~(\ref{eq:spin1XY})) or families of Hamiltonians with $\eta$-pairing states of Eq.~(\ref{eq:HubbardQP}))~\cite{mark2020eta}.
\subsubsection{Miscellaneous Examples}

\begin{figure}
    \centering
    \begin{tabular}{cc}
    \includegraphics[scale=0.22]{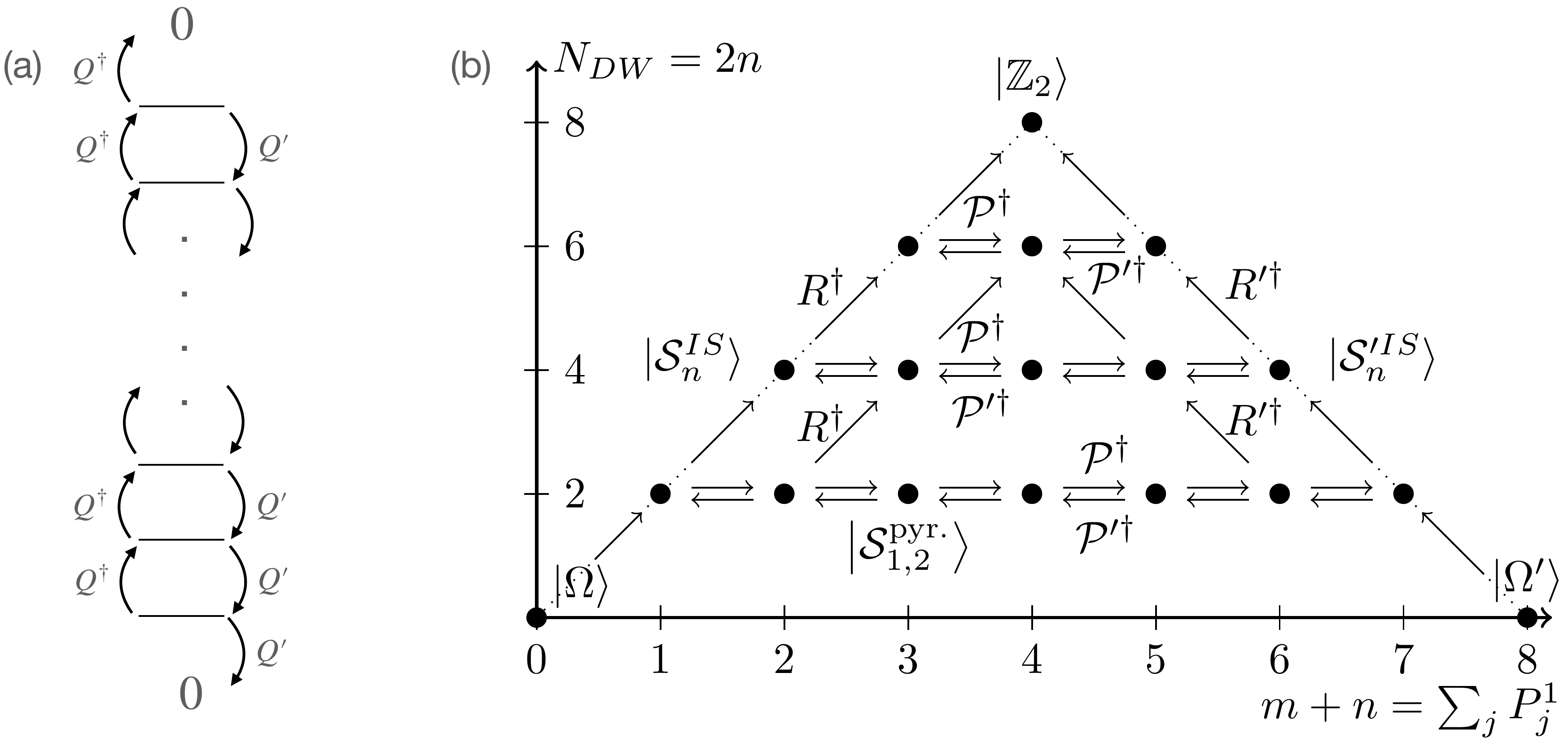}
    \end{tabular}
    \caption{Various structures of QMBS. (a) Equally spaced tower of QMBS with a raising operator $Q^\dagger$ and a lowering operator $Q'$, which might or might not be the Hermitian conjugate of $Q^\dagger$.
    The raising operator is usually a sum of single-site operators, although there are examples where it is a sum of multi-site operators or completely non-local. 
    $Q^\dagger$ and $Q'$ annihilate the highest and lowest states of the tower respectively. 
    (b) Figure reproduced from Ref.~\cite{mark2020unified} showing a pyramid of QMBS in a certain domain-wall conserving model. 
    Each dot represents a QMBS eigenstate labelled by certain quantum numbers  $N_{DW}$ and $\sum_j{P^1_j}$.
    The various QMBS eigenstates are related ladder operators such as $\mP^\dagger$, $\mP'^\dagger$, $R^\dagger$, etc. 
    This shows the wide variety of structures of QMBS that can occur in physically relevant Hamiltonians. 
    }
    \label{fig:towerpyramid}
\end{figure}
All of the examples of QMBS discussed above consist of towers of single-site quasiparticles, i.e., the states are obtained by the multiple action of a quasiparticle creation operator that is a sum of on-site terms.
A different class of models with towers of QMBS are those with raising operators that are multi-site quasiparticles, i.e., the quasiparticle creation operator is a sum of multi-site but local terms.
These include the second tower of the spin-1 XY model discovered in Ref.~\cite{schecter2019weak} and subsequently studied in Ref.~\cite{chattopadhyay2019quantum}, where the origin of towers of eigenstates was traced to the existence of ``virtual entanglement pairs".
Large classes of models with multi-site quasiparticle QMBS were constructed based on the Onsager algebra in Ref.~\cite{shibata2020onsager, vanvoorden2021disorder}, and also systematically constructed from parent Hamiltonians of MPS in Ref.~\cite{moudgalya2020large}.
Finally, we note that equally spaced eigenstates and revivals in non-integrable models are known to appear in several other systems and lattices~\cite{hudomal2020quantum, lee2020exact, hart2020compact, chertkov2021motif}, sometimes without a quasiparticle structure of the eigenstates.  
In addition, examples of QMBS where the ``raising operator" $Q^\dagger$ is non-local were constructed in the context of systems with quantum group symmetries~\cite{odea2020from}, as well as in the DWC model~\cite{mark2020unified}.
A tower of QMBS created by a non-local operator also appears in models constructed to embed ``rainbow states" in the spectrum~\cite{langlett2021rainbow}, which, unlike most examples of QMBS, obey a volume-law scaling of EE under most choices of the bipartition while still violating ETH.
Another example of a tower of QMBS and associated revivals with volume-law scaling of EE was shown to exist in a certain multi-component system~\cite{zhao2021orthogonal}. 
Apart from simple towers of states, Refs.~\cite{mark2020unified, odea2020from} construct examples of ``pyramids" of QMBS created by the actions of multiple raising and lowering operators on a simple eigenstate, e.g., as shown in Fig.~\ref{fig:towerpyramid}b.
\subsection{Entanglement of quasiparticle towers of states}\label{subsec:QPentanglement}
\begin{figure}
    \centering
    \includegraphics[scale=0.25]{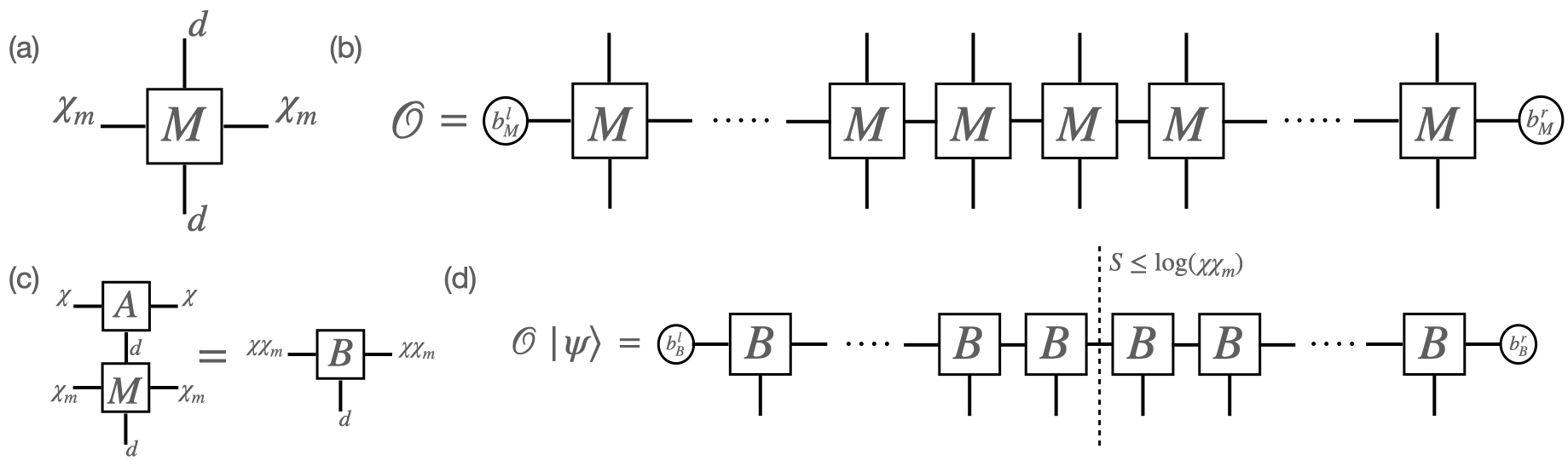}
    \caption{(a) $d \times d \times \chi_m \times \chi_m$ tensor representing an MPO. $d$ is the physical dimension, and $\chi_m$ is the bond dimension (b) Operator $\mO$ represented in MPO form (c) Action of an MPO $M$ with bond dimension $\chi_m$ on an MPS $A$ with bond dimension $\chi$ gives an MPS $B$ with bond dimension $\chi \chi_m$ (d) MPS representation of the state $\mO \ket{\psi}$. The entanglement entropy of the state $\mO\ket{\psi}$ is bounded by $\log\left(\chi\chi_m\right)$. }
    \label{fig:MPO}
\end{figure}
We now turn to the EE of states composed of multiple quasiparticles on a fixed background MPS, which form the towers of QMBS eigenstates in several models in Secs.~\ref{subsec:SGAtowers} and \ref{subsec:othertowers}. 
For pedagogical reasons, we restrict ourselves to one-dimensional systems and states of the form
\begin{equation}
    \ket{\psi_n} = (Q^\dagger)^n \ket{\psi_0},\;\; Q^\dagger = \sumal{j}{}{e^{i k j} q^\dagger_j},
\label{eq:psinform}
\end{equation}
where $q^\dagger_j$ is a single-site operator and $\ket{\psi_0}$ is a MPS. 
These properties hold for the raising operators in several models with towers of QMBS, including the $\eta^\dagger$ and $Q^\dagger$ operators in the Hubbard and AKLT models respectively.
A simple way to bound the entanglement entropy of states $\{\ket{\psi_n}\}$ is to study its MPS form, which can be derived using the Matrix Product Operator (MPO) form~\cite{pirvu2010matrix,schollwock2011density,mcculloch2007density,verstraete2008matrix,moudgalya2018b} of the creation operators $(Q^\dagger)^n$ and the MPS form for $\ket{\psi_0}$. 
Simple translation invariant operators $\mathcal{O}$ typically admit an exact MPO representation in terms of a $d \times d \times \chi_m \times \chi_m$ tensors $M$ shown in Fig.~\ref{fig:MPO}a.
Such an operator $\mathcal{O}$ is depicted diagrammatically in Fig.~\ref{fig:MPO}b.
$d$ and $\chi_m$ are referred to as physical and bond dimensions of the MPO respectively, and the diagrammatic notation is analogous to that for MPS, shown in Fig.~\ref{fig:MPS}b and discussed in Sec.~\ref{subsec:entanglement}.
A state defined by the action of an MPO on an MPS has a natural MPS description in terms of the tensor $B$ depicted diagrammatically in Fig.~\ref{fig:MPO}c.
$B$ is sometimes referred to as an MPO $\times$ MPS~\cite{moudgalya2018b}, and it has a bond dimension of $\chi_m \chi$, where $\chi_m$ and $\chi$ are the bond dimensions of the MPO and MPS respectively. 
Hence, according to Eq.~(\ref{eq:MPSEEbound}), the EE of the state $B$ represents is bounded by $S \leq \log\left(\chi_m \chi\right)$ (see Fig.~\ref{fig:MPO}d).  
In the following, we use this bound to provide a simple proof for the sub-volume law EE scaling of the towers of QMBS eigenstates discussed in Secs.~\ref{subsec:SGAtowers} and \ref{subsec:othertowers}.
Using standard methods~\cite{crosswhite2008fsa, motruk2016density, moudgalya2018b}, we can construct an efficient MPO of bond dimension $\chi_m = n+1$ for the operator $(Q^\dagger)^n$ ($n$ being an integer). 
Although a general expression is complicated (see Appendix A of Ref.~\cite{odea2020from}),\footnote{The elements $M_{\alpha,\beta}$ of the MPO tensor and the boundary vectors $(b^l_M)_\alpha$ and $(b^r_M)_\alpha$ in the general case read
\begin{equation*}
M_{\alpha,\beta} = e^{i(n-\alpha+1)k} (q^\dagger)^{\beta-\alpha} \times \frac{(n!)^{\frac{\beta-\alpha}{n}}}{(\beta-\alpha)!}\delta_{\beta \geq \alpha},\;\;\;(b^l_M)_\alpha = \delta_{\alpha,1},\;\;\;(b^r_M)_\alpha = \delta_{\alpha, n +1}
\end{equation*}}
in the simple case when $(q^\dagger_j)^2 = 0$ and $k = \pi$, which happens in several QMBS models including the Hubbard and AKLT model, the MPO tensor and the boundary vectors read~\cite{moudgalya2018b}
\begin{equation}
        M =     
        \begin{pmatrix}
        (-1)^n \mathds{1} & (-1)^n q^\dagger & 0 & \dots & 0\\
        0 & (-1)^{n-1} \mathds{1} & (-1)^{n-1} q^\dagger & \ddots & \vdots \\
        \vdots & \ddots & \ddots & \ddots & 0 \\
        \vdots & \ddots & \ddots & -\mathds{1} & -q^\dagger \\
        0 & \dots & \dots & 0 & \mathds{1} \\
    \end{pmatrix},\;\;\begin{array}{c}
         {b^l_M}^T = \begin{pmatrix} 1 & 0 & \cdots & 0\end{pmatrix}  \\
         {b^r_M}^T = \begin{pmatrix} 0 & \cdots & 0& 1\end{pmatrix} 
    \end{array},
\label{eq:towerMPO}
\end{equation}
If the bond dimension of the MPS representation of $\ket{\psi_0}$ is $D$, the state $\ket{\psi_n}$ has an MPS representation with bond dimension $\chi = D (n +1)$. 
This establishes an upper bound on the EE of the states of the tower to grow as $S \leq \log [D(n+1)]$.
For $D$ that is independent of system size (since $\ket{\psi_0}$ is typically the ground state), and for a state $\ket{\psi_n}$ with an extensive number of quasiparticles (meaning $n \propto L$), the EE thus grows with system size $L$ as $S \sim \log L$, a sub-volume scaling. 
These results are consistent with the sub-volume law scaling found in states with multiple identical quasiparticles on top of a product state, which has been studied in a variety of settings~\cite{vafek2017entanglement, castro2018entanglement, schecter2019weak}, including the scaling depicted in the inset of Fig.~\ref{fig:scarproperties_ent}.
Further, exact results for the EE of certain quasiparticle eigenstates that appear as QMBS in various systems, including those where the raising operators do not exactly obey the precise properties of Eq.~(\ref{eq:psinform}), have been obtained in Refs.~\cite{vafek2017entanglement, schecter2019weak, chattopadhyay2019quantum, odea2020from}, and they all follow a similar sub-volume law scaling. 
Indeed, MPOs with bond dimension $\chi_m \propto n$ can also be obtained for operators $(Q^\dagger)^n$  when the quasiparticle creation operators $q^\dagger_j$ have supports over multiple sites or sometimes also when they are non-local~\cite{odea2020from}, although writing out their explicit form can be tedious.
These results show that towers of QMBS exhibit an EE scaling that is inconsistent with ETH predictions.
On a different note, the MPS forms of the QMBS can also be used to identify ``topological" properties such as projective representations of the MPS or degeneracies in their entanglement spectrum, as shown for the AKLT model in Ref.~\cite{moudgalya2018b}. 
\subsection{Revivals from simple initial states}\label{subsec:revivals}
\begin{figure}
    \centering
    \includegraphics[scale=0.4]{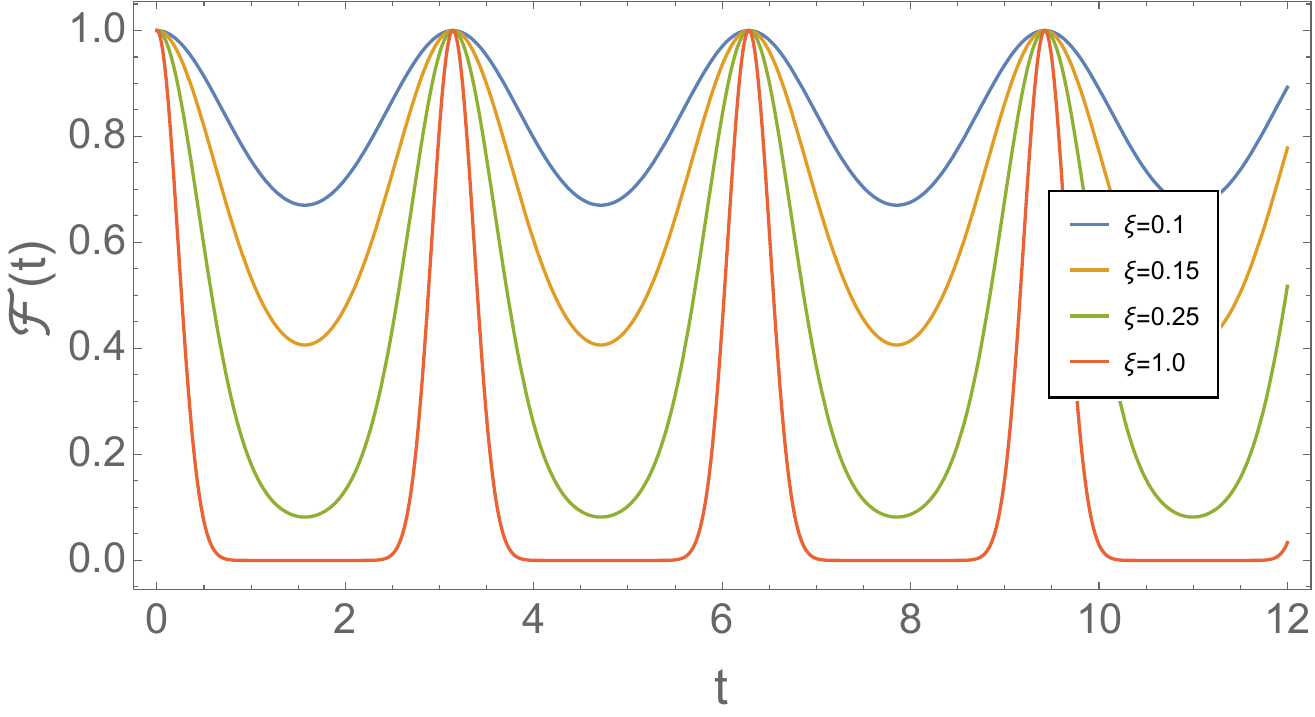}
    \caption{(Color online) Fidelity of initial states of the form of $\ket{\psi_{\textrm{in}}(\xi)}$ of Eq.~(\ref{eq:specialinitialstates}) for the one-dimensional spin-1 XY Hamiltonian of Eq.~(\ref{eq:spin1XY}) with OBC and $h = 1$ and system size $L = 10$.
     The revivals occur due to the equal spacing of the QMBS eigenstates,  hence the nature of the plot does not depend on the precise scarred Hamiltonian.
     The time period of the revivals is $2\pi/\mE$,  where $\mE$ is the energy spacing between the states of the tower. 
    }
    \label{fig:scarproperties_dyn}
\end{figure}
Given a quasiparticle tower of equally-spaced QMBS of the form $\{(\Qd)^n \ket{\psi_0}\}$ with $0 \leq n \leq N-1$ such that $(\Qd)^N\ket{\psi_0} = 0$, it is natural to ask what kinds of simple initial states can be constructed within the QMBS subspace.
The first example of such a construction was demonstrated in Ref.~\cite{iadecola2020quantum} in the context of the DWC model mentioned in Sec.~\ref{subsec:othertowers}. 
We illustrate this construction, focusing on one-dimensional systems where the quasiparticle creation operator $\Qd$ has the form $\Qd = \sum_{j = 1}^{L}{e^{ikj}q^\dagger_j}$, where $q^\dagger_j$ is a strictly local operator in the vicinity of site $j$, possibly with support over multiple sites. 
For simplicity, we further assume that the quasiparticle creation operators on different sites commute, i.e. $[q^\dagger_l, q^\dagger_m] = 0$, a feature that is true for several examples of QMBS.
A simple family of initial states that is in the scarred subspace is then given by~\cite{iadecola2020quantum, mark2020eta, ren2021deformed}
\begin{equation}
    \ket{\psi_{\textrm{in}}(\xi)} = \prodal{j = 1}{L}{\exp(\xi e^{ikj} q^\dagger_j)}\ket{\psi_0} = \exp(\xi Q^\dagger)\ket{\psi_0} = \sumal{n = 0}{N}{\frac{\xi^n}{n!} (\Qd)^n \ket{\psi_0}},
\label{eq:specialinitialstates}
\end{equation}
where we have excluded the normalization factor that depends on the precise details of $q^\dagger_j$.
Note that in several common examples of QMBS discussed in Sec.~\ref{sec:towers}, $k = \pi$ and $(q^\dagger_j)^2 = 0$, hence $\exp(\xi e^{i k j} q^\dagger_j) = (1 + (-1)^j \xi q^\dagger_j)$.
The fidelity $\mathcal{F}(t)$ of Eq.~(\ref{eq:fidelity}) can be directly computed starting from the states of Eq.~(\ref{eq:specialinitialstates}).
If the QMBS states $\{(Q^\dagger)^n \ket{\psi_0}\}$ have energies $\{E_0 + n\mE\}$ under a scarred Hamiltonian $H_{\scar}$, the expression for the fidelity (after including the normalization factor) reads
\begin{equation}
    \mF(t) = \frac{|\bra{\psi_{\textrm{in}}(\xi)} e^{-i H_{\scar} t} \ket{\psi_{\textrm{in}}(\xi)}|^2}{|\braket{\psi_{\textrm{in}}(\xi)}{\psi_{\textrm{in}}(\xi)}|^2} = \frac{|\sumal{n = 0}{N}{\frac{\xi^{2n} e^{-i n \mathcal{E}t}}{(n!)^2}\mathcal{N}_n}|^2}{|\sumal{n = 0}{N}{\frac{\xi^{2n}}{(n!)^2}\mathcal{N}_n}|^2},\;\;\; \mathcal{N}_n \equiv \bra{\psi_0}(Q)^n (Q^\dagger)^n\ket{\psi_0},  
\label{eq:scarfidelity}
\end{equation}
where $\mathcal{N}_n$'s are the normalization factor of the QMBS eigenstate $(Q^\dagger)^n\ket{\psi_0}$. 
While the normalization factors $\mathcal{N}_n$ might be in general hard to compute analytically in general, they can be explicitly evaluated in some cases, e.g., in the spin-1 XY model~\cite{iadecola2019quantum} or the Hubbard model~\cite{vafek2017entanglement}, which leads to dramatic simplifications in Eq.~(\ref{eq:scarfidelity}).
Using the expression of Eq.~(\ref{eq:scarfidelity}),  it is easy to see that $\mF(t + \frac{2\pi}{\mE}) = \mF(t)$, hence fidelity of the initial states $\ket{\psi_{\textrm{in}}(\xi)}$ exhibits revivals with a time period given by $2\pi/\mE$.
In Fig.~\ref{fig:scarproperties_dyn}, we show the fidelity revivals for various values of $\xi$ for the spin-1 XY model.
This periodicity is in stark contrast to the fidelity of a random state with the same energy expectation value as $\ket{\psi_{\textrm{in}}(\xi)}$, which typically quickly decays to a value exponentially small in system size.
Since $\ket{\psi_{in}(\xi)}$ is constructed by the action of $L$ one-site or two-site (depending on the number of sites $q^\dagger_j$ has support on) operators $\{\exp(\xi e^{ikj} q^\dagger_j)\}$ on $\ket{\psi_0}$, the bond-dimension of its MPS representation is $\mathcal{O}(1)$ more than the bond-dimension of the MPS for $\ket{\psi_0}$.
Hence, if $\ket{\psi_0}$ obeys area-law entanglement (i.e., if it admits an MPS representation of finite bond dimension), the family of states $\ket{\psi_{\textrm{in}}(\xi)}$ also obeys area-law entanglement, even though some of the states $(Q^\dagger)^n\ket{\psi_0}$ are not area-law entangled as discussed in Sec.~\ref{subsec:QPentanglement}. 
Due to the MPS structure of $\ket{\psi_{\textrm{in}}(\xi)}$, we can construct its local parent Hamiltonian~\cite{perezgarcia2007matrix, moudgalya2020large} $H_{\textrm{in}}(\xi)$ for which $\ket{\psi_{\textrm{in}}(\xi)}$ is an exact ground state, although not necessarily unique. 
An explicit construction of such a family of gapped parent Hamiltonians $H_{\textrm{in}}(\xi)$ is provided in Ref.~\cite{iadecola2020quantum}, where they turn out to be closely related to the Rokhsar-Kivelson type Hamiltonians studied in Ref.~\cite{lesanovsky2011many}. 
This construction of initial states within the QMBS subspace that are gapped ground states of different Hamiltonians is highly desirable, since it provides an experimentally feasible method to build initial states that show anomalous dynamics. 
\section{Unified Formalisms}\label{sec:unifiedformalism}
Given the large number of models exhibiting QMBS, there have been several attempts to unify them into systematic formalisms~\cite{mori2017eth, mark2020unified, moudgalya2020eta, pakrouski2020many, ren2020quasisymmetry, odea2020from}.
In spite of these works, it is not clear to date whether any of these are exhaustive, and the precise relations between these formalisms are yet to be carefully worked out.
Nevertheless, we now give a broad overview of the different approaches to unify QMBS, which roughly fall into three categories. 
Throughout this section, we use $H_{\scar}$ to denote the Hamiltonian of interest, i.e., the one with QMBS eigenstates, $\mT$ to denote the subspace spanned by the QMBS eigenstates, and $Q^\dagger$ to refer to the raising operator.
In particular, we will frequently refer back to the example of the Hubbard model discussed in Sec.~\ref{subsec:SGAtowers}, where $H_{\scar}$ is given by Eq.~(\ref{eq:Int}).
For convenience of illustration, we rewrite $H_{\scar}$ here and split it into three parts
\begin{gather}
H_{\scar} = \widehat{T} + \widehat{U} + \widehat{V},\;\;\;\hT \equiv  -\sumal{\langle \vr, \vr'\rangle}{}{t_{\vr, \vr'} \hT_{\vr, \vr'}} = -\sumal{\langle \vr, \vr'\rangle}{}{t_{\vr, \vr'} \sumal{\sigma \in \{\uparrow, \downarrow\}}{}{\left(\cd_{\vr, \sigma} c_{\vr', \sigma} + h.c.\right)}}\nn \\
\hU \equiv \sumal{\vr}{}{\left(U \hn_{\vr, \uparrow} \hn_{\vr, \downarrow} - \mu\sumal{\sigma \in \{\uparrow, \downarrow\}}{}{\hn_{\vr, \sigma}}\right)},\;\;\;\hV \equiv \sumal{\dlangle \vr, \vr'\drangle}{}{V_{\vr,\vr'} S^z_{\vr} S^z_{\vr'}},
\label{eq:HInt}
\end{gather}
where we have used the same notation as Eq.~(\ref{eq:Int}).
The subspace $\mT$ and the raising operator $Q^\dagger$ corresponding to this model are given by
\begin{equation}
    Q^\dagger \equiv \eta^\dagger = \sumal{\vr}{}{e^{i\bp\cdot\vr} \cd_{\vr, \uparrow} \cd_{\vr, \downarrow}},\;\;\;\mT \equiv \textrm{span}\{\ket{\Omega}, \eta^\dagger \ket{\Omega}, \cdots, (\eta^\dagger)^L\ket{\Omega}\}.
\label{eq:Hparams}
\end{equation}
\subsection{Shiraishi-Mori embedding formalism}\label{subsec:SMformalism}
The first systematic method of ``embedding" exact eigenstates into the spectrum of non-integrable Hamiltonians was introduced by Shiraishi and Mori (SM) in Ref.~\cite{mori2017eth}.  
The SM formalism uses a set of strictly local (generically multi-site) projectors $\{P_i\}$ that need not commute with each other, and a target space $\mT$ defined as the common subspace of states annihilated by all the projectors, i.e. $\mT = \{\ket{\psi} : P_i\ket{\psi} = 0\;\;\forall i\}$.
Given a target space $\mT$, any term $H_0$ that commutes with all of the $P_i$'s leaves the target space invariant (i.e, $H_0\ket{\psi} \in \mT$ if $\ket{\psi} \in \mT$ since $P_i H_0\ket{\psi} = H_0 P_i\ket{\psi} = 0$).
Hence, $H_0$ can be diagonalized within $\mT$, and the corresponding eigenstates are the eigenstates of any Hamiltonian $H_{\scar}$ of the form
\begin{equation}
    H_{\scar} = \sumal{i}{}{P_i h_i P_i} + H_0,\;\;\;[H_0, P_i] = 0\;\;\;\forall i. 
\label{eq:shiraishimori}
\end{equation}
where $h_i$ is an arbitrary local operator.
For generic choices of $h_i$, $H_{\scar}$ is non-integrable, and the states in $\mT$ are eigenstates in the middle of the spectrum. 
Provided the states have sub-volume-law EE scaling, they are violations of strong ETH of the Hamiltonian $H_{\scar}$, and thus examples of QMBS of $H_{\scar}$. 
While the original examples in Ref.~\cite{mori2017eth} only included ``isolated" QMBS that are not equally spaced towers of states (we discuss these in Sec.~\ref{sec:isolated}), it was later realized that towers of QMBS in several models can also be captured by this formalism.
Examples include the QMBS towers in the spin-1 XY model (see Appendix C of Ref.~\cite{schecter2019weak}), $\eta$-pairing in the Hubbard model~\cite{mark2020eta} and $H_{\scar}$ of Eq.~(\ref{eq:HInt}), the toy model studied in Ref.~\cite{choi2018emergent}, although recasting these Hamiltonians in the form of Eq.~(\ref{eq:shiraishimori}) can be tedious.
\subsection{SGA-based formalism}\label{subsec:SGAbased}
\subsubsection{MLM Framework}
A different unified framework was introduced by Mark-Lin-Motrunich (MLM) in Ref.~\cite{mark2020unified}, generalizing the idea of SGAs discussed in Sec.~\ref{subsec:SGAtowers}.
They consider a manifold of states $\mW$, and impose the following SGA condition restricted to $\mW$
\begin{equation}
    [H_{\scar}, Q^\dagger]\mW = \mE Q^\dagger \mW,\;\;\; Q^\dagger \mW \subseteq \mW, 
\label{eq:manifoldSGA}
\end{equation}
where $Q^\dagger$ is the QMBS tower creation operator, and $\hO \mW$ is an abuse of notation that refers to the subspace obtained by the action of an operator $\hO$ onto states in $\mW$.
Note that when $\mW$ is the full Hilbert space, Eq.~(\ref{eq:manifoldSGA}) reduces to the the SGA condition of Eq.~(\ref{eq:SGA}).
Consequently, given an eigenstate $\ket{\psi_0}$ of $H_{\scar}$ within the subspace $\mW$, we obtain a tower of eigenstates of $H_{\scar}$ of the form $\{(Q^\dagger)^n \ket{\psi_0}\}$.
However, note that unless $\mW$ is the full Hilbert space, Eq.~(\ref{eq:manifoldSGA}) does not imply that the lowering operator of the tower is the Hermitian conjugate of $Q^\dagger$; hence in general it can be a different operator $Q'$, as shown in Fig.~\ref{fig:towerpyramid}a.
For example, for the deformations of the Hubbard model such as $H_{\scar}$ of Eq.~(\ref{eq:HInt}), the subspace $\mW$ is equal to the subspace $\mT$ spanned by the tower of states in Eq.~(\ref{eq:Hparams}).
Similarly, this formalism was demonstrated to capture all the QMBS in the spin-1 XY model of Eq.~(\ref{eq:spin1XY}), all the spin-$S$ $SO(3)$-symmetric AKLT models, and the DWC model discussed in Sec.~\ref{sec:towers}.
Insights from the MLM formalism also led to the discovery of large families of nearest-neighbor Hamiltonians that share the same QMBS eigenstates as these models.
Note that although the subspace $\mW$ is the same as the QMBS subspace $\mT$ in simple examples, it was sometimes observed to be larger than the subspace $\mT$, for example in the case of the AKLT model~\cite{mark2020unified}.
\subsubsection{RSGA formalism}
A closely related formalism was independently introduced in Ref.~\cite{moudgalya2020eta}, where, instead of working with manifolds of states $\mW$, sufficient conditions for the existence of towers of eigenstates of the form $\{(Q^\dagger)^n\ket{\psi_0}\}$ were provided in terms of the state $\ket{\psi_0}$, the Hamiltonian $H_{\scar}$, and the operator $Q^\dagger$.
In particular, defining $H_0 \equiv H_{\scar},\;\; H_{n+1} \equiv [H_n, Q^\dagger],\;\;\forall n \geq 0$, examples of QMBS were said to exhibit a Restricted Spectrum Generating Algebra of order $M$ (RSGA-$M$) if the following conditions are satisfied 
\begin{eqnarray}
    &&\textrm{(i)}\ H_{\scar}\ket{\psi_0} = E_0\ket{\psi_0}, \;\;\; \textrm{(ii)}\ H_1\ket{\psi_0} = \mE Q^\dagger\ket{\psi_0}\nn \\
    &&\textrm{(iii)}\ H_n\ket{\psi_0} = 0 \;\; \forall\ n,\ 2 \leq n \leq M,\;\;\;\textrm{(iv)}\ \twopartdef{H_n \neq 0}{n \leq M}{H_n = 0}{n = M+1}.
\label{eq:RSGA2cond}
\end{eqnarray}
Explicit examples of QMBS in Hubbard-like models exhibiting an RSGA-$M$ for any $M$ were constructed in Ref.~\cite{moudgalya2020eta}. 
For example, $H_{\scar}$ of Eq.~(\ref{eq:Int}) can be shown to satisfy Eq.~(\ref{eq:RSGA2cond}) with $\ket{\psi_0} = \ket{\Omega}$, $\mE = (U - 2\mu)$, $Q^\dagger = \eta^\dagger$, $E_0 = 0$ for $M = 1$, hence it falls into the category of RSGA-1.
The RSGA formalism provides a finer classification of QMBS that are part of the MLM formalism.
For example, the spin-1 XY model exhibits an RSGA-1, while the spin-1 AKLT model exhibits an RSGA-2. 
Recently, Ref.~\cite{tang2021multimagnon} connected the conditions of Eq.~(\ref{eq:RSGA2cond}) to the properties of spherical tensor operators, which were then used to systematically construct families of QMBS Hamiltonians satisfying the RSGA conditions. 
Note that similar to the MLM formalism, Eq.~(\ref{eq:RSGA2cond}) does not restrict the form of the lowering operator of the tower to be the Hermitian conjugate of $Q^\dagger$; hence in general it can be a different operator $Q'$, as shown in Fig.~\ref{fig:towerpyramid}a.
\subsection{Symmetry-based formalisms}\label{subsec:symmetrybased}
\begin{figure}
    \centering
    \includegraphics[scale=0.25]{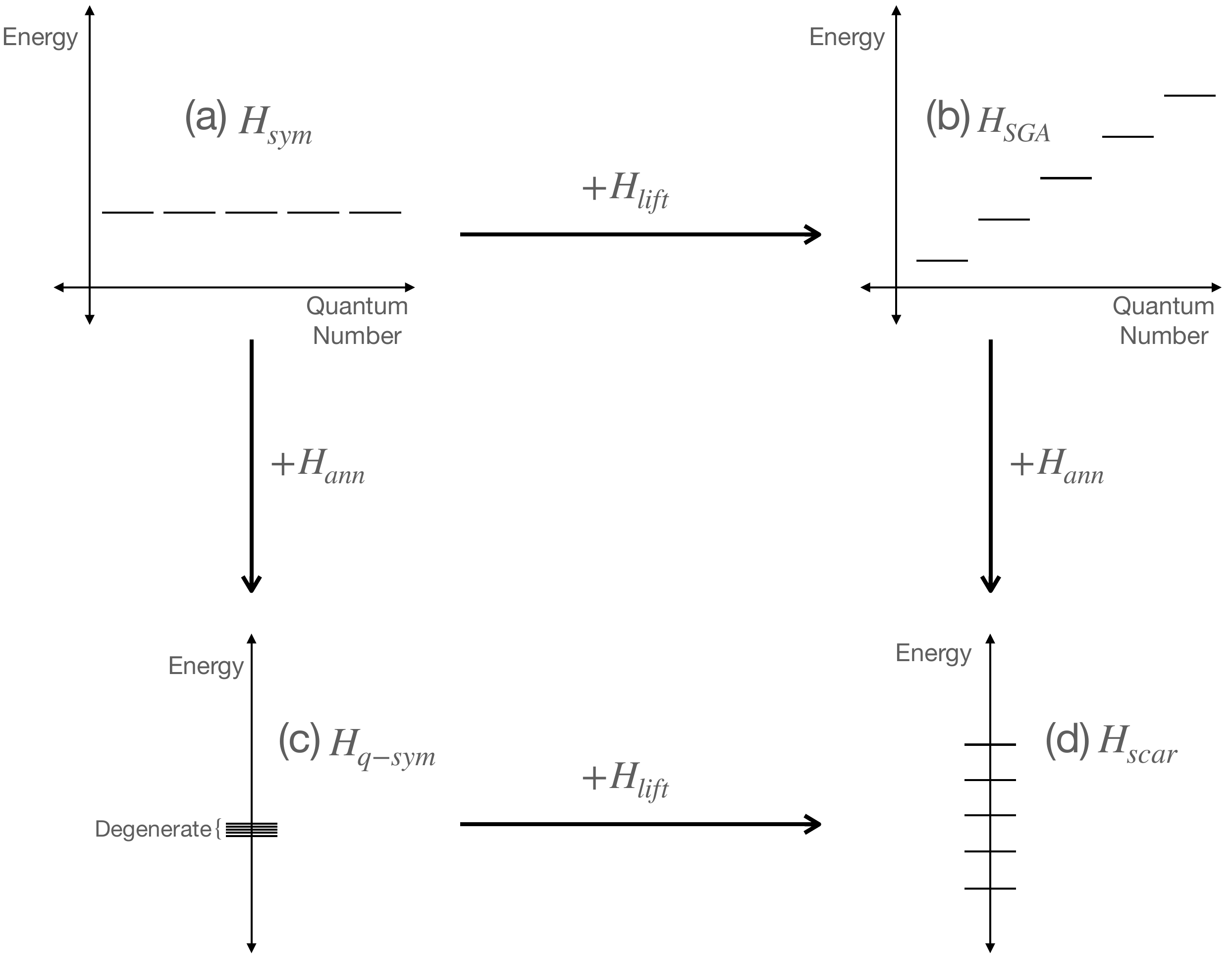}
    \caption{Schematic overview of the symmetry-based unified formalisms for QMBS, discussed in Sec.~\ref{subsec:symmetrybased}. The figure only shows the eigenstates of interest, which eventually form the QMBS of $H_{\scar}$.
    (a) Spectrum of a symmetric Hamiltonian $H_{\sym}$. The degeneracy can either be viewed to be a consequence of the non-Abelian symmetry (TT formalism) or due to the fact that they are singlets of the group generated by terms of $H_{\sym}$ (GI formalism). The eigenstates are uniquely specified by certain symmetry quantum numbers of $H_{\sym}$. 
    (b) Spectrum of the Hamiltonian $H_{\SGA} = H_{\sym} + H_{\lift}$ that appears in the TT formalism. While the degeneracy of these eigenstates is lifted, the eigenstates are still uniquely specified by certain symmetry quantum numbers of the $H_{\SGA}$. 
    (c) Spectrum of the Hamiltonian $H_{\qsym} = H_{\sym} + H_{\ann}$ that appears in the QS and GI formalisms, showing a manifold of exactly degenerate eigenstates.  
    The symmetry of $H_{\sym}$ is broken by the addition of $H_{\ann}$, but the degeneracy of the eigenstates is preserved. 
    (d) Spectrum of the Hamiltonian $H_{\scar} = H_{\qsym} + H_{\lift} = H_{\SGA} + H_{\ann}$. 
    Here the eigenstates are referred to as towers of QMBS, since they are typically in the middle of the spectrum, and are not distinguished by any symmetry quantum number of $H_{\scar}$. 
    }
    \label{fig:symmetrybased}
\end{figure}
The SGA-based formalisms discussed in Sec.~\ref{subsec:SGAbased} were subsequently extended in Refs.~\cite{ren2020quasisymmetry, odea2020from, pakrouski2020many}.
We refer to these as symmetry-based formalisms since the starting points for the construction of $H_{\scar}$ are highly symmetric Hamiltonians for which the the QMBS are degenerate eigenstates, and the raising and lowering operators $Q$ and $Q^\dagger$ are symmetries.
The search for such symmetric Hamiltonians was already initiated in Ref.~\cite{mark2020eta}, which employed similar ingredients as the symmetry-based formalisms we discuss below, along with the MLM framework to systematically search for Hamiltonians with QMBS.
The symmetry-based formalisms are summarized in Fig.~\ref{fig:symmetrybased}, and can be broadly classified into two categories. 
\subsubsection{Quasisymmetry and Tunnels to Towers Formalisms}
To construct the appropriate symmetric Hamiltonian, Refs.~\cite{ren2020quasisymmetry, odea2020from} focus on the symmetry algebra generated by the raising and lowering operators $Q$ and $Q^\dagger$, which is $SU(2)$ in the case of $\eta$-pairing discussed in Sec.~\ref{subsec:SGAtowers} (see Eq.~(\ref{eq:su2syms})).
In Ref.~\cite{ren2020quasisymmetry}, the Hamiltonian $H_{\scar}$ exhibiting towers of QMBS is decomposed into two parts  
\begin{equation}
    H_{\scar} = H_{\textrm{q-sym}} + H_{\lift},
\label{eq:QSHamil}
\end{equation}
where the QMBS eigenstates are degenerate in $H_{\textrm{q-sym}}$, and $H_{\lift}$ lifts their degeneracy into an equally spaced tower of states, analogous to a magnetic field (see Figs.~\ref{fig:symmetrybased}c and \ref{fig:symmetrybased}d). 
The degeneracy of the QMBS eigenstates in $H_{\textrm{q-sym}}$ is shown to be a consequence of their invariance under a  ``Quasisymmetry" (QS), which is defined to be a usual on-site symmetry restricted to a particular subspace of the full Hilbert space.
In particular, given any group element $g$ of a symmetry group $G$ with the on-site unitary action $\widehat{u}(g)$, a subspace $\mW$ is referred to as having a quasisymmetry if it is invariant under the action of the symmetry group, i.e., if $\widehat{u}(g) \mW \subseteq \mW$ for all $g \in G$. 
For example, in the Hubbard model, the subspace $\mT$ of Eq.~(\ref{eq:Hparams}) is said to possess an $SU(2)$ quasisymmetry, since it is invariant under the action of the pseudospin $SU(2)$ unitary operators $\widehat{u}(\vec{\alpha}) = \exp(i \vec{\alpha}\cdot\vec{\eta})$.
Ref.~\cite{ren2020quasisymmetry} provides a systematic method, similar to the Shiraishi-Mori construction of Sec.~\ref{subsec:SMformalism}, to construct Hamiltonians $H_{\textrm{q-sym}}$ with the desired quasisymmetric subspace as a degenerate eigenspace.
In fact, in the cases where the SM formalism and the QS formalisms overlap, the first term in Eq.~(\ref{eq:shiraishimori}) is an example of $H_{\qsym}$.
For example, given the quasisymmetric subspace $\mT$ of Eq.~(\ref{eq:Hparams}), one such Hamiltonian is $H_{\textrm{q-sym}} = \hT + \hV$.
Further, in the case where the quasisymmetry is a Lie group, Ref.~\cite{ren2020quasisymmetry} shows that there are natural candidates for $H_{\lift}$ that lifts the degeneracy of the quasisymmetric subspace in $H_{\textrm{q-sym}}$.
In the case of the subspace $\mT$, one such Hamiltonian is $H_{\lift} = \hU$, which lifts the degeneracy into equally spaced towers of states with $\mE = (U - 2\mu)$.
In all, this provides a systematic approach to construct Hamiltonians with towers of QMBS, which have the form of Eq.~(\ref{eq:QSHamil})
As illustrated in Ref.~\cite{ren2020quasisymmetry}, this quasisymmetry formalism captures several examples of QMBS in the literature.
Further, the notion of quasisymmetries and the associated formalism were recently generalized to include many additional examples of QMBS~\cite{ren2021deformed}.
On the other hand, Ref.~\cite{odea2020from} exemplified a ``Tunnels to Towers" (TT) approach to systematically construct models with towers of states, where Hamiltonians $H_{\scar}$ exhibiting QMBS is decomposed into three parts 
\begin{equation}
    H_{\scar} = H_{\sym} + H_{\lift} + H_{\ann}.
\label{eq:TTHamil}
\end{equation}
$H_{\sym}$ is a Hamiltonian with a conventional non-Abelian symmetry such as $SU(2)$ that protects the degeneracy of a multiplet of eigenstates that eventually become the tower of QMBS of $H_{\scar}$ (see Fig.~\ref{fig:symmetrybased}a). 
In this construction, the eigenstates that are part of the chosen multiplet are uniquely specified by their quantum numbers under symmetries of $H_{\sym}$, and are related by the actions of raising and lowering operators associated with the non-Abelian symmetry (e.g. by $S^+$ and $S^-$ in the case of $SU(2)$)
$H_{\lift}$ is a term that can be systematically added to lift the degeneracy of these eigenstates of $H_{\sym}$ into an equally spaced tower (see Fig.~\ref{fig:symmetrybased}b), such that the resulting Hamiltonian, given by
\begin{equation}
    H_{\SGA} = H_{\sym} + H_{\lift},
\label{eq:SGAHamil}
\end{equation}
exhibits an SGA property of Eq.~(\ref{eq:SGA}).
Nevertheless, these eigenstates are still not referred to as examples of QMBS of $H_{\SGA}$ since they can still be uniquely specified by quantum numbers under symmetries of $H_{\SGA}$ (see Fig.~\ref{fig:symmetrybased}b).
Finally $H_{\ann}$ contains terms that annihilate the tower of states and can be systematically added to break the SGA property of $H_{\SGA}$ while preserving (typically annihilating) the tower of states as eigenstates, arriving at $H_{\scar}$ (see Figs.~\ref{fig:symmetrybased}b and \ref{fig:symmetrybased}d). 
For example, in the case of $H_{\scar}$ of Eq.~(\ref{eq:HInt}), the symmetric Hamiltonian is $H_{\sym} = \hT$, which possess the pseudospin $SU(2)$ symmetry, and the multiplet of degenerate eigenstates (with the degeneracy protected by the pseudospin $SU(2)$ symmetry) is the subspace $\mT$ of Eq.~(\ref{eq:Hparams}).
Further, the term that lifts the degeneracy of this multiplet of states is $H_{\lift} = \hU$.
$H_{\SGA}$ is thus the usual Hubbard model $\hhub$ of Eq.~(\ref{eq:FermiHubbard}), which exhibits an SGA of Eq.~(\ref{eq:HubbardSGA}), as discussed in Sec.~\ref{subsec:SGAtowers}.
Finally, the term that breaks the SGA property while annihilating the eigenstates in $\mT$ is $H_{\ann} = \hV$, and thus the decomposition of $H_{\scar}$ in Eq.~(\ref{eq:HInt}) is directly the TT decomposition.
As evident from the example of $H_{\scar}$ of Eq.~(\ref{eq:HInt}), the QS and TT constructions of Refs.~\cite{ren2020quasisymmetry, odea2020from} reproduce the SGA-based construction of scars discussed in Refs.~\cite{mark2020eta, moudgalya2020eta} when the symmetry group is restricted to $SU(2)$. 
Further, these formalisms are also closely related to each other, and in the cases they overlap, the $H_{\textrm{q-sym}}$ in the QS formalism can be decomposed as (see Figs.~\ref{fig:symmetrybased}a and \ref{fig:symmetrybased}c)
\begin{equation}
    H_{\textrm{q-sym}} = H_{\sym} + H_{\ann},
\label{eq:QSTTrelation}
\end{equation}
and can be interpreted in the language of the TT formalism and vice-versa.
Apart from towers of states, both mechanisms were also demonstrated for $SU(3)$ groups, where two independent raising operators lead to ``pyramids" of QMBS instead of towers.
However, they differ in certain aspects.
While the QS formalism applies only to groups protecting the degeneracy in $H_{\sym}$, the TT mechanism was also demonstrated for $SU(2)_q$, a quantum group protecting the degeneracy, in which case the raising and lowering operators for the towers of states are non-local in nature.
On the other hand, for the TT formalism to work, it requires $H_{\sym}$ to already possess analytically tractable eigenstates, which is a built-in feature in the QS formalism.    
\subsubsection{Group Invariant Formalism}
A complementary understanding of towers of QMBS was provided in a Group-Invariant (GI) formalism introduced in Ref.~\cite{pakrouski2020many}.
The starting points are symmetric Hamiltonians $H_{\sym}$ that are quadratic fermionic hopping terms of the form $T_A = \sum_{\vr, \vr', \sigma}{A_{\vr, \vr'} \cd_{\vr, \sigma} c_{\vr',\sigma}}$ on $L$ sites where $A$ is a Hermitian matrix, and the indices $\vr$ and $\sigma$ label the lattice site and the spin respectively. 
Instead of studying the symmetry group of such Hamiltonians, Ref.~\cite{pakrouski2020many} focuses on the algebra generated by individual terms in $H_{\sym}$.
For example, for $H_{\scar}$ in Eq.~(\ref{eq:HInt}), the corresponding $H_{\sym}$ is $\hT$, and the object of interest in the GI formalism is the algebra generated by the individual nearest-neighbor hopping terms $\{\hT_{\vr, \vr'}\}$.
Such quadratic hopping terms $\{T_A\}$ are shown to be the generators of a Lie group, e.g., $SO(L)$ in the case of $\{\hT_{\vr, \vr'}\}$.
The associated symmetry group (i.e., the group of all unitary operators that commute with all these quadratic terms $\{T_A\}$) is a different Lie group, e.g., $SO(4) \approx (SU(2) \times SU(2))/Z_2$ in the case of $\{\hT_{\vr, \vr'}\}$, which is usually referred to as the ``symmetry group" of the system.  
While QS and TT formalisms use the properties of the latter symmetry group to construct Hamiltonians with towers of QMBS, the GI formalism constructs Hamiltonians with QMBS using the properties of the former group (e.g., $SO(L)$).
In particular, they show that several examples of QMBS states are ``invariant" under the action of the Lie group generated by the quadratic terms.
These states are referred to as one-dimensional representations or ``singlets" of corresponding Lie groups.
For example, the QMBS eigenstates in $\mT$ of Eq.~(\ref{eq:Hparams}) are all annihilated by individual hopping terms $\{\hT_{\vr, \vr'}\}$, and hence are invariant under the action of the $SO(L)$ group generated by these terms (i.e., the states are referred to as singlets of $SO(L)$). 
Hence the degeneracy of the QMBS eigenstates under $H_{\sym}$ shown in Fig.~\ref{fig:symmetrybased}a is viewed to be a consequence of the states being annihilated by the terms of $H_{\sym}$ (i.e., they are singlets of the group generated by the terms of $H_{\sym}$). 
Using the fact that the singlets are annihilated by the quadratic terms, Ref.~\cite{pakrouski2020many} provides a systematic construction for $H_{\scar}$, similar to the Shiraishi-Mori construction of Sec.~\ref{subsec:SMformalism}, which can be written in the form
\begin{equation}
    H_{\scar} = \sumal{A}{}{O_A T_A} + H_{\lift},
\label{eq:GIHamil}
\end{equation}
where $\{O_A\}$ are arbitrary operators, the summation runs over some set of matrices $A$, and $H_{\lift}$ is some Hamiltonian that leaves the singlet space (i.e., the QMBS space) invariant, which guarantees that singlets of the group generated by $\{T_A\}$ are generically non-degenerate eigenstates of $H_{\scar}$.
Further, note that the first term in Eq.~(\ref{eq:GIHamil}) is equivalent to $H_{\textrm{q-sym}}$ in the QS construction (see Fig.~\ref{fig:symmetrybased}c) and also to the first term in the SM formalism of Eq.~(\ref{eq:shiraishimori}).
Note that Eq.~(\ref{eq:GIHamil}) can also be related to the TT construction by decomposing the first term in Eq.~(\ref{eq:GIHamil})  according to Eq.~(\ref{eq:QSTTrelation}).
An advantage of the GI formalism is that it reveals large symmetries of the QMBS eigenstates that are not evident in the QS and TT formalisms.
For example, since the states in $\mT$ of Eq.~(\ref{eq:Hparams}) are invariant under the action of the group $SO(L)$, they can be viewed as being $SO(L)$-symmetric, which in particular implies that they are invariant under a permutation of the sites of the lattice (since the permutation group $S_L$ is a subgroup of $SO(L)$). 
Note that this formalism was recently extended to include additional examples of QMBS in fermionic Hamiltonians~\cite{pakrouski2021group}.
\section{Isolated QMBS}\label{sec:isolated}
\begin{figure}
    \centering
    \includegraphics[scale=0.25]{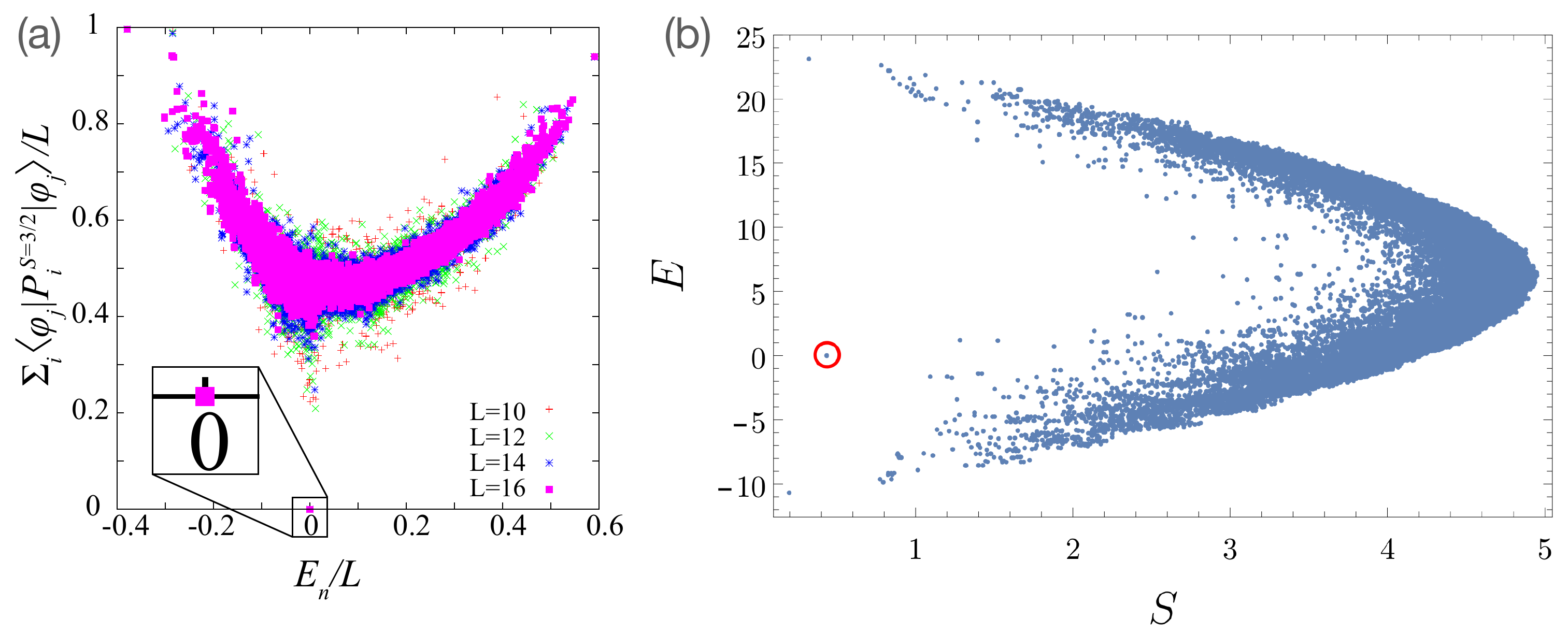}
    \caption{(Color online) Examples of Isolated QMBS.  (a) Figure reproduced from Ref.~\cite{mori2017eth} shows the ETH violation in an isolated QMBS via the expectation value of a certain local operator.  The inset shows the expectation value in an isolated QMBS, which strongly deviates from that of typical states in the spectrum.  (b) Figure reproduced from Ref.~\cite{ok2019topological} shows the EE of the eigenstates as a function of energy in a certain Hamiltonian exhibiting isolated QMBS that show topological order. The circled point is the QMBS eigenstate.}
    \label{fig:isolatedQMBS}
\end{figure}

While QMBS are commonly associated with revivals and the existence of towers of equally-spaced eigenstates in the spectrum, several examples of QMBS that do not involve a exactly solvable tower of states.
Such examples consist of any number of states embedded in the middle of the spectrum, ranging from an $\mathcal{O}(1)$ number to exponentially many.
Some examples of isolated QMBS are shown in Fig.~\ref{fig:isolatedQMBS}.
\subsection{Survey of isolated QMBS}\label{subsec:SMformalismisolated}
The Shiraishi-Mori (SM) formalism discussed in Sec.~\ref{subsec:SMformalism} was introduced in Ref.~\cite{mori2017eth} to systematically embed QMBS eigenstates in the middle of the spectrum of non-integrable Hamiltonians of the form of Eq.~(\ref{eq:shiraishimori}). 
This formalism was explicitly demonstrated in Ref.~\cite{mori2017eth} by constructing such a non-integrable Hamiltonian with two Majumdar-Ghosh Hamiltonian ground states~\cite{majumdarghosh1969} as QMBS eigenstates in the middle of its spectrum.
This construction utilized the frustration-free property of the Majumdar-Ghosh Hamiltonian ground states, i.e., they are annihilated by individual three-site projectors $P_i$'s, and as discussed in Sec.~\ref{subsec:SMformalism}, they form the target subspace $\mT$. 
This procedure can be applied to other Valence Bond States including AKLT state, which have the property that they are annihilated by an appropriate choice of local operators $P_i$~\cite{aklt1987rigorous, majumdarghosh1969}.
In fact, a larger class of states can be a part of the target space $\mT$.
For example, in the case of states with MPS representation in one-dimension (or PEPS in higher dimensions), the appropriate $P_i$'s can be derived using the parent Hamiltonian construction~\cite{perezgarcia2007matrix, moudgalya2020large} or via Eigenstate-to-Hamiltonian construction algorithms~\cite{qi2019determininglocal, chertkov2018computational}.
Other examples of isolated QMBS in the literature include Hamiltonians with QMBS possessing (symmetry-protected) topological order~\cite{ok2019topological, srivatsa2020quantum, jeyaretnam2021quantum}, certain QMBS that appear in frustrated models~\cite{lee2020exact, mcclarty2020disorder, chertkov2021motif}, ``dimerized" states in lattice models with supersymmetry in arbitrary dimensions~\cite{surace2020weakergodicity}, exact localized states in Hubbard-like models~\cite{iadecola2018exact}, and certain eigenstates in transverse field Ising ladders~\cite{vanvoorden2021disorder} or Abelian lattice gauge theory on small ladders~\cite{banerjee2021quantum}.
We note that many of these examples of isolated QMBS appear to be special cases of the SM formalism, although the connection is not always immediately obvious. 
Finally, we note that Ref.~\cite{mori2017eth} also illustrated an example of embedding where the subspace $\mT$ has an exponentially large dimension (while being a measure-zero set in the thermodynamic limit).
Although eigenstates within that subspace are not solvable, they nevertheless violate the conventional form of ETH~\cite{mori2017eth, shiraishimorireply}.
However, the subspace can also be thought of as an exponentially large ``Krylov subspace" distinguished by a non-local symmetry~\cite{mondaini2018comment, moudgalya2021hilbert}, which makes it closer to examples of fragmentation that we will discuss in Sec.~\ref{sec:fragmentation}.
\subsection{PXP models}\label{subsec:PXPmodels}
An important class of isolated exact QMBS eigenstates appear in the context of PXP models, an effective model for the dynamics of Rydberg atoms~\cite{lesanovsky2011many, turner2017quantum} on arbitrary lattices or graphs.
As discussed in Sec.~\ref{sec:intro}, the experimental realization and the approximate QMBS in the PXP model played an important role in the emergence of QMBS as a field. 
For convenience and due to its importance, we briefly describe the system here and refer readers to the reviews on the subject for more details~\cite{SerbynReview, papic2021weak}.
\subsubsection{Model and Approximate QMBS}
In a certain limit where the nearest neighbor van-der-Walls interaction between Rydberg atoms is much larger than the detuning and the Rabi frequency, the Rydberg atoms can be modelled by two-level systems, either in its ground state or excited state, which we denote by $\downarrow$ and $\uparrow$ respectively~\cite{lesanovsky2012interacting, turner2017quantum}. 
Furthermore, in this limit, the interactions between excited Rydberg atoms effectively forbid the simultaneous excitation of nearest-neighbor atoms, hence the effective low-energy Hilbert space consists of all configurations without nearest neighbor excitations (i.e. configurations of the form $\ket{\cdots \uparrow \uparrow \cdots}$ are absent), which is sometimes referred to as the Fibonacci Hilbert space~\cite{lesanovsky2012interacting}.
The effective Hamiltonian $H_{\textrm{PXP}}$ within this low-energy subspace allows Rydberg atoms to freely transition between their ground and excited states provided neighboring atoms are excited~\cite{lesanovsky2012interacting}.
Its expression reads (with PBC)
\begin{equation}
    H_{\textrm{PXP}} = \twopartdef{\sum_j{P_{j-1} X_j P_{j+1}}}{\textrm{in one dimension}}{\sum_j{\left(X_j \prod_{i \in N(j)}{P_i}\right)}}{\textrm{on arbitrary lattices}},\;\;\
\label{eq:PXPdefn}
\end{equation}
where $j$ runs over the sites of the lattice, $P_j = \left(\ket{\downarrow}\bra{\downarrow}\right)_j$ is a projector onto the ground state of the atom on site $j$ that and $N(j)$ denotes the set of neighbors of site $j$.
Note that the PXP Hamiltonian of Eq.~(\ref{eq:PXPdefn}) cannot create any nearest-neighbor excitations, and hence preserves the Fibonacci Hilbert space.
The PXP model in any dimension admits a particle-hole symmetry generated by the operator $\prod_j{Z_j}$~\cite{turner2018quantum}, where $Z_j$ is the Pauli-$Z$ matrix acting on site $j$, hence their energy spectra are symmetric around $E = 0$.  
In addition, on lattices on which the PXP model has an inversion symmetry (e.g. the 1d PXP model), exhibit an exponentially large manifold of zero energy ($E = 0$) eigenstates that can be shown to be a consequence of the inversion and particle-hole symmetries~\cite{turner2017quantum, turner2018quantum, schecter2018many, moudgalya2020quantum}. 
Note that the 1d PXP model also appears as an effective Hamiltonian in a variety of different contexts, including Fibonacci anyon chains~\cite{trebst2008fibonacci, lesanovsky2012interacting, chandran2020absence}, Ising models on dimer ladders~\cite{moessner2001ising, laumann2012quantum}, lattice gauge theories~\cite{surace2020lattice, chen2021emergent}, as well as in models with dipole moment conservation~\cite{fendley2004competing,sala2019ergodicity}, in particular the quantum Hall effect on a thin torus~\cite{moudgalya2020quantum}. 
The 1d PXP model has been observed to host an approximate tower of QMBS that leads to anomalously long revivals~\cite{bernien2017probing,turner2017quantum, turner2018quantum}.
These approximate QMBS in the PXP model have been studied using a wide variety of techniques that yield several insights into their origin~\cite{khemani2019int,ho2018periodic, choi2018emergent, michailidis2020slow, bull2019systematic, turner2021correspondence, bull2020quantum}, which are reviewed in more detail in Refs.~\cite{SerbynReview, papic2021weak}. 
However, we emphasize that this tower of QMBS in the PXP model~\cite{turner2017quantum} or in its deformations with almost perfect revivals~\cite{choi2018emergent} are not examples of exact QMBS, and are hence beyond the scope of this review.
\subsubsection{Exact QMBS in 1d}
Beyond the tower of approximate QMBS, exact eigenstates of the 1d PXP model with area-law entanglement and simple MPS expressions were constructed in Ref.~\cite{lin2019exact}, which are examples of exact QMBS.
Two exact states with $E = 0$ were obtained for OBC and PBC, whereas two additional exact states with $E = \pm \sqrt{2}$ were obtained for OBC, which differ from the $E = 0$ exact states at the boundaries. 
The OBC eigenstates at non-zero energy are certainly in the middle of the energy spectrum and should be considered as examples of QMBS.
The case is slightly different for the exact states with $E = 0$, which are examples of QMBS only if typical eigenstates in the exponentially large zero energy state manifold satisfy ETH, numerical evidence for which was found in Ref.~\cite{schecter2018many}.
These exact eigenstates also appear to be base states for variational quasiparticle constructions of the approximate towers of QMBS in the PXP model~\cite{lin2019exact}, complementary to other approximations for the tower of QMBS in the PXP models~\cite{iadecola2019quantum, mark2019new}. 
More examples of such exact eigenstates in PXP-like models were obtained in Refs.~\cite{surace2021exactstability, karle2021arealaw}.
Furthermore, numerical observations in Ref.~\cite{karle2021arealaw} also suggest that some low-entanglement states similar to the exact states in the PXP model exist within the exponentially large manifold of $E = 0$ eigenstates of \textit{all} local Hamiltonians with inversion and particle-hole symmetries, although they might not have an MPS form with finite bond-dimension. 
Finally, we note that Ref.~\cite{shiraishi2019connection} connected these exact states in the PXP model to the Shiraishi-Mori formalism discussed in Sec.~\ref{subsec:SMformalism}.
\subsubsection{Exact QMBS in Higher Dimensions}
PXP models in higher dimensions also admit exact QMBS eigenstates.
Ref.~\cite{lin2020quantum} constructed exponentially many exact $E = 0$ ``dimerized" eigenstates (similar to the ones in Ref.~\cite{surace2020lattice}) in the 2d PXP model on square and rotated-square lattices, which can also be generalized to the 3d PXP model on cubic lattices.
As demonstrated there, these eigenstates also have direct implications for dynamics of states on Rydberg arrays.
However, since they all have zero energy, they are different from the approximate towers of equally-spaced eigenstates that lead to revivals observed in the deformed 2d PXP model studied in Ref.~\cite{michailidis2020stabilizing}.
\subsection{Other exact eigenstates}
To complete our discussion on exact results on excited states, we briefly survey some examples of exact excited states that are not considered examples of QMBS due to their position in the energy spectrum (i.e., they are typically not in the bulk of the spectrum). 
However, in many cases we expect that these eigenstates are QMBS of appropriately modified Hamiltonians where such states are ``embedded" into the middle of the spectrum following ideas similar to the Shiraishi-Mori formalism discussed in Sec.~\ref{subsec:SMformalism}.
One class of exact excited states are single quasiparticle excited states above a frustration-free ground state of a Hamiltonian, which have an area-law entanglement~\cite{moudgalya2018b, pizorn2012universality}.
Examples of such eigenstates close to the edges of the spectrum date back to early works in the Majumdar-Ghosh model~\cite{caspers1982some, caspers1984majumdar}, as well as two exact low-energy excited ``Arovas" states in the spin-1 AKLT model~\cite{arovas1989two}, which were later generalized to any integer spin-$S$~\cite{moudgalya2018a}.
More recently, Refs.~\cite{lesanovsky2011many, mark2019new} solved for quasiparticle exact excited states in a frustration-free Hamiltonian modelling Rydberg interactions similar to the PXP model, and Ref.~\cite{mark2019new} also used the nature of these states to obtain better variational expressions for the approximate towers of QMBS in the PXP model. 
Similarly, exact expressions for several families of one or two quasiparticle eigenstates close to the ferromagnetic state of certain $SU(2)$-symmetric models have been obtained.
The simplest of these are the well-known spin-waves on top of the ferromagnet, and more involved examples of such states can be found in the spin-1 AKLT model~\cite{moudgalya2018a}.
Finally, we note that examples of quasiparticle states also exist in the literature in higher dimensions, particularly in the $U(1)$-symmetric XY model (also known as the XX model) on hypercubic lattices~\cite{bibikov2017bethe}, and in the Hubbard model on square lattices~\cite{ren2018slater, ye2018exact}. 
These eigenstates have $\mathcal{O}(L)$ quasiparticles in an $L \times L$ system, and the quasiparticle number density (and hence their energy density)  goes to zero in the thermodynamic limit.
Hence we do not expect them to be direct examples of QMBS in their respective models. 
\section{Hilbert Space Fragmentation and Krylov Subspaces}\label{sec:fragmentation}
Hilbert space fragmentation is a related phenomenon of ergodicity breaking.
In this section, we discuss this phenomenon, as well as its similarities and differences with QMBS.
Given a quantum system with Hilbert space $\mH$ and Hamiltonian $H$, we can decompose the Hilbert space into dynamically disconnected subspaces, referred to as \textit{Krylov subspaces} (or sometimes as \textit{fragments} or \textit{subsectors}), as follows
\begin{equation}
    \mH = \bigoplus_{j = 1}^K{\mK_j},\;\;\;\mK_j = \textrm{span}_t\{e^{-i H t}\ket{\psi_j}\},  
\label{eq:Hilbertsplitting}
\end{equation}
where $\textrm{span}_t\{e^{-i H t}\ket{\psi_j}\} \equiv \textrm{span}\{\ket{\psi_j}, H\ket{\psi_j}, H^2\ket{\psi_j}, \cdots,\}$ denotes the subspace spanned by time-evolution of the state $\ket{\psi_j}$, and $K$ is the number of Krylov subspaces. 
Note that the $\ket{\psi_j}$'s in Eq.~(\ref{eq:Hilbertsplitting}) are chosen such that the subspaces $\mK_j$'s are distinct. 
As discussed in Sec.~\ref{subsec:ergodicitybreaking}, Hilbert space fragmentation referred to the phenomenon where the system possesses exponentially many Krylov subspaces, i.e. $K \sim \exp(L)$ for a system of size $L$. 
The decomposition of Eq.~(\ref{eq:Hilbertsplitting}) is trivial if $\ket{\psi_j}$'s are eigenstates of $H$, and is expected if the Hamiltonian $H$ possesses certain symmetries such that different $\ket{\psi_j}$'s in Eq.~(\ref{eq:Hilbertsplitting}) have different symmetry quantum numbers. 
However, the decomposition for fragmented systems is different in two ways.
First, $\ket{\psi_j}$'s are typically chosen to be product states, usually motivated by their more straightforward experimental preparation. 
For a non-integrable Hamiltonian, such a choice usually leads the subspace $\mK_j$ to be generically the \textit{full} Hilbert space after resolving symmetries of the system.
However, the different Krylov subspaces $\mK_j$ in fragmented systems are not distinguished by quantum numbers corresponding to any obvious local symmetries of $H$.
Second, for generic systems with conventional symmetries such as $\mathbb{Z}_2$, $U(1)$, or $SU(2)$, the number of Krylov subspaces $K$ either stays constant or grows polynomially with increasing system size, whereas it grows exponentially in fragmented systems~\cite{sala2019ergodicity,khemani2019local,moudgalya2021hilbert}.
Furthermore, the dimensions of the Krylov subspaces in fragmented systems typically have a wide range, from one-dimensional ``frozen" product states where all terms of Hamiltonian act trivially, to subspaces with exponentially large dimension that can be studied in terms of a Krylov-restricted effective Hamiltonian.
Fragmentation was explicitly pointed out in the context of dipole-moment or center-of-mass conserving systems~\cite{pai2018localization, sala2019ergodicity, khemani2019local, moudgalya2019thermalization}, although similar phenomena have been discussed in several works~\cite{ritort2003glassy, bergholtz2005half, bergholtz2006one, olmos2010thermalization, sikora2011extended, nakamura2012exactly, gopalakrishnan2018facilitated, lan2018quantum, moudgalya2020quantum}.
In Sec.~\ref{subsec:dipoleconservation} we discuss a simple example of Hilbert space fragmentation in a dipole-moment conserving system, which illustrates several essential features of fragmented systems.
In Secs.~\ref{subsec:miscfrag}, we survey some other models in the literature that demonstrate fragmentation and comment on some general features of fragmented systems. 
In Sec.~\ref{subsec:fragmentationdynamics}, we discuss the implications of fragmentation to dynamics and connections to QMBS.
Finally, we dedicate Sec.~\ref{subsec:autocorr} to the discussion of autocorrelation functions, an important diagnostic in the context of fragmentation, and discuss its properties.    
\subsection{Simple example: Pair-hopping model}\label{subsec:dipoleconservation}
We start with a simple example of Hilbert space fragmentation that occurs in an interacting model of electrons that conserves the total dipole moment of the system in addition to the total charge. 
Systems conserving dipole moment or center-of-mass were first discussed in Ref.~\cite{seidel2005incompressible}, in the quest to build featureless Mott insulators.
They naturally arise in physical systems in two different contexts.
First, in quantum Hall effect on a thin cylinder, where the effective Hamiltonian with translation invariant interactions projected onto a single Landau level~\cite{bergholtz2006one, bergholtz2008quantum, moudgalya2020quantum, lee2015geometric, papic2014solvable,rezayi1994,nakamura2012exactly} exhibits dipole moment conservation. 
Second, they also appear as effective Hamiltonians within resonant subspaces in the interacting Wannier-Stark problem, i.e., interacting fermions hopping on a finite one-dimensional lattice, subject to a strong electric field~\cite{van2018bloch,schulz2019stark, moudgalya2019thermalization, taylor2019experimental}.
Such a system has been probed in many recent experiments~\cite{guardado2020subdiffusion, scherg2020observing, kohlert2021experimental}, and Hilbert space fragmentation is believed to contribute to the observed slow dynamics~\cite{taylor2019experimental, doggen2021stark}.

\subsubsection{Model and symmetries}
We now illustrate Hilbert space fragmentation in dipole-moment conserving systems with the help of a simple one-dimensional spinless fermionic ``pair-hopping model"~\cite{seidel2005incompressible, moudgalya2020quantum, moudgalya2019thermalization} $\hph$ with OBC, which is the ``quantum part" (i.e., neglecting electrostatic terms) of the pseudopotential Hamiltonian in the $\nu = 1/3$ Fractional Quantum Hall effect~\cite{lee2015geometric} in the thin torus limit~\cite{lee2015geometric, moudgalya2020quantum}:
\begin{equation}
\label{eq:pairhopping}
\hph = \sum_{j = 1}^{L-3}{H_j} = \sum_{j = 1}^{L-3}{\left(\cd_j \cd_{j+3} c_{j + 2} c_{j + 1} + h.c.\right)}.
\end{equation}
The terms $H_j$ implement the transitions $\ket{1 \spa 0 \spa  0 \spa 1} \leftrightarrow \ket{0 \spa 1 \spa 1 \spa 0}$, where $\ket{a \spa b \spa c \spa d}$ denote the occupancies of four consecutive sites on the chain, and the model preserves the dipole moment (i.e. center-of-mass position)~\cite{seidel2005incompressible}, given by the operator 
\begin{equation}
    \widehat{D} \equiv \sum_{j = 1}^{L}{j \hat{n}_j},
\end{equation}
where $\hat{n}_j$ is the fermion number operator on site $j$.
Given the set of allowed transitions by the terms $\{H_j\}$, we can study its dynamically disconnected Krylov subspaces $\{\mK_j\}$ of Eq.~(\ref{eq:Hilbertsplitting}).
We are only interested in the dynamics of initial product states, which are more easily accessible to experiments, and hence we consider Krylov subspaces $\mK_j$ generated by product states $\ket{\psi_j}$. 
Although we will only consider the translation invariant model of Eq.~(\ref{eq:pairhopping}), note that the Krylov subspaces we study only depend on the structure of the transitions of $\{H_j\}$, hence the Krylov subspaces of the entire family of models of the form $\sum_j{J_j H_j}$ for arbitrary couplings $J_j$ are the same. 
\subsubsection{Frozen configurations and small Krylov subspaces}
Exponentially many of these Krylov subspaces are one-dimensional \textit{frozen} configurations---product states that are eigenstates of $H$. 
This is a direct consequence of the ``sparsity" of transitions that the terms $\{H_j\}$ implement, i.e. the Hamiltonian vanishes on any product state not containing the patterns $``\cdots 0110 \cdots"$ or $``\cdots 1001\cdots"$ on four consecutive sites.
Since there are \textit{exponentially} many such patterns, there are equally many one-dimensional Krylov subspaces.
Further, Krylov subspaces can be constructed by embedding finite \textit{active} blocks, i.e. regions where the Hamiltonian acts non-trivially, into any frozen configuration, thereby leading to exponentially many Krylov subspaces with dimension of $\mathcal{O}(1)$~\cite{sala2019ergodicity,khemani2019local}.
For example, the following configurations $\ket{\psi_\pm}$
\begin{eqnarray}
    &\ket{\psi_\pm} = \frac{1}{\sqrt{2}}\left(\ket{111000\cdots111000\fbox{1001}111000\cdots111000}\right. \nn \\
    &\left.\pm \ket{111000\cdots111000\fbox{0110}111000\cdots111000}\right)  
\end{eqnarray}
are composed of one active block (boxed) sandwiched between frozen configurations, and they span a two-dimensional Krylov subspace. 
The presence of exponentially many frozen states and states with frozen regions within each symmetry sector in the Hilbert space has direct implications to the dynamics of such systems: time-evolution starting from a random product state looks very different from the behavior expected for typical thermal non-integrable models.
\subsubsection{Exponentially large Krylov subspaces}
Apart from frozen configurations and Krylov subspaces of small dimension, the pair-hopping model also exhibits Krylov subspaces with exponentially large dimensions that scale with system-size as $\sim \alpha^L$ as $L \to \infty$ and $1 < \alpha < 2$.
As discussed in Ref.~\cite{moudgalya2019thermalization}, these subspaces can be highly non-locally constrained, and certain Krylov subspaces in $\hph$ are characterized by a certain non-local string order. 
Furthermore, the properties of these subspaces can be vastly different, demonstrating the rich dynamical structure inherent to systems with Hilbert space fragmentation. 
For example, the Hamiltonian restricted to a given subspace can be either integrable or non-integrable (either satisfying a restricted form of ETH or MBL), and subspaces of different types can coexist.
We now provide one example of an integrable Krylov subspace with exponentially large dimension in the pair-hopping model $\hph$ of Eq.~(\ref{eq:pairhopping}) via a spin mapping demonstrated in Refs.~\cite{bergholtz2005half, bergholtz2006one, bergholtz2008quantum, moudgalya2019thermalization}.  
The Hamiltonian $\hph$ with even system size $L = 2N$ has an additional symmetry: sublattice particle number conservation~\cite{moudgalya2019thermalization}, and we group sites $2j - 1$, $2j$ of the original lattice into a new site $j$ so as to form a new chain with $N = L/2$ sites.
It is convenient to work in terms of new degrees of freedom for these composite sites defined as $\ket{\uparrow} \equiv \ket{0 \spa 1}$ $\ket{\downarrow} \equiv \ket{1 \spa 0}$.
The action of the terms $H_j$ of Eq.~\eqref{eq:pairhopping}, when written in terms of the composite spins, is simply given by $\ket{\ \fbox{01}\ \fbox{10}\ } \leftrightarrow \ket{\ \fbox{10}\ \fbox{01}\ } \iff \ket{\uparrow \downarrow} \leftrightarrow \ket{\downarrow\uparrow}$. 
Hence for any Krylov subspace generated by a product state $\ket{\psi_0}$ with only composite spin degrees of freedom $\uparrow$ and $\downarrow$,  the action of the Hamiltonian restricted to the Krylov subspace interchanges the spins, and hence exactly maps onto a spin-1/2 XX model:
\begin{equation}
    H_{XX}\left[N\right] \equiv \sum_{j = 1}^{N}{\left(\sigma^+_j \sigma^-_{j+1} + \sigma^-_j \sigma^+_{j+1}\right)} \, ,
\label{eq:hamilXX}
\end{equation}
where $\{\sigma^+_j\}$ and $\{\sigma^-_j\}$ are Pauli matrices on site $j$. 
As is well known, the Hamiltonian Eq.~(\ref{eq:hamilXX}) can be solved using a Jordan-Wigner transformation~\cite{lieb1963exact}, upon which it maps onto a non-interacting fermion problem. 
However, as can be readily shown, these are \textit{not} the only states within the same charge and dipole moment sector, providing evidence for Hilbert space fragmentation in the pair-hopping Hamiltonian $\hph$. 
\subsubsection{Krylov subspaces due to blockades}
Additional integrable or non-integrable Krylov subspaces can be systematically constructed by inserting \textit{blockades} in the system, i.e., frozen configurations on a part of the system that remain unchanged under the action of the Hamiltonian. 
For example, consider Krylov subspaces generated by product states of the $\ket{\cdots 111 \cdots}$ or $\ket{\cdots 000 \cdots}$, where $\cdots$ denotes active regions where the terms $\{H_j\}$ of $\hph$ act non-trivially.
Configurations such as $111$ or $000$ embedded in the middle of the chain do not change under the action of the local terms $H_j$.
Hence they can be used to separate active regions of the chain, leading to exponentially many new Krylov subspaces. 
The Hamiltonian restricted to such Krylov subspaces with multiple active regions separated by blockades is simply the sum of commuting terms that act on different active regions of the chain.  
\subsubsection{Strong v/s weak fragmentation}
In addition to the ```minimal-range" dipole conserving model of Eq.~(\ref{eq:pairhopping}), we can introduce longer-range dipole moment conserving terms such as $\left(\cd_j \cd_{j + 5} c_{j+1} c_{j+4} + h.c.\right)$, which connect several of the Krylov subspaces of the minimal-range model~\cite{sala2019ergodicity, moudgalya2019thermalization}. 
Nevertheless, Refs.~\cite{sala2019ergodicity, khemani2019local} showed that Hamiltonians with dipole moment conservation and terms of any finite range is guaranteed to exhibit Hilbert space fragmentation.
A simple proof relies on building exponentially many frozen patterns that are annihilated by dipole moment conserving terms of any finite range.
For example, states of the form $\ket{0 \cdots 0 1 \cdots 1 0 \cdots 0 1 \cdots 1}$, where $0 \cdots 0$ and $1 \cdots 1$ are clusters of at least $m$ sites with identical occupation, are frozen under the action of any dipole conserving term acting on less than or equal to $m$ consecutive sites. 
However, as discussed in Ref.~\cite{sala2019ergodicity}, adding longer range terms to the minimal dipole conserving model changes the nature of fragmentation in the system from ``strong" to ``weak", which are defined as follows.
Referring to the dimension of the largest Krylov subspace as $D_{\textrm{max}} = \textrm{max}_j\{\textrm{dim}(\mK_j)\}$, and full Hilbert space dimension (after resolving all conventional symmetries) as $D$, Ref.~\cite{sala2019ergodicity} classified fragmented systems into two types: \emph{strongly fragmented} when $D_{\max}/D \rightarrow 0$ and \emph{weakly fragmented} when $D_{\max}/D \rightarrow 1$ respectively in the thermodynamic limit. 
Note that these notions of strong and weak fragmentation only apply \text{within} usual (e.g., charge and dipole) symmetry sectors~\cite{moudgalya2021hilbert}.
Indeed, Ref.~\cite{morningstar2020kinetically} provided an example of a dipole moment conserving Hamiltonian that exhibits strong fragmentation within certain quantum number sectors while exhibiting weak fragmentation in other sectors.
\subsection{Survey of other examples in the literature}\label{subsec:miscfrag}
We now survey several other examples of Hilbert space fragmentation in the literature, including those that do not involve dipole moment conserving models.
\subsubsection{Spin-1 Dipole Conserving Model}
Apart from dipole conserving systems of spinless fermions that we have discussed in Sec.~\ref{subsec:dipoleconservation}, spin-1 dipole conserving model, given by the Hamiltonian 
\begin{equation}
H_3 \equiv \sum_j{(S^-_{j-1} (S^+_j)^2 S^-_{j+1} + h.c.)},
\label{eq:H3hamil}
\end{equation}
where $S^{\pm}_j$'s are the spin-1 raising and lowering operators on site $j$. 
This Hamiltonian, and its Floquet version, have been studied in detail in Refs.~\cite{pai2018localization,sala2019ergodicity, khemani2019local, rakovszky2019statistical}. 
They show that the model possesses several similar features as $\hph$, i.e., it hosts exponentially many frozen eigenstates and Krylov subspaces, most of which feature blockades, as well as exponentially large non-locally constrained Krylov subspaces.
As shown in Ref.~\cite{rakovszky2019statistical}, the model also exhibits integrable subspaces that map onto spin-1/2 XX models.
While the minimal spin-1 dipole conserving model exhibits strong fragmentation in all charge and dipole quantum number sectors, longer range spin-1 dipole-conserving model were numerically observed to exhibit weak fragmentation~\cite{sala2019ergodicity} in the largest quantum number sector. 
However, Ref.~\cite{morningstar2020kinetically} observed a ``freezing transition" from weak to strong fragmentation as a function of filling factor (i.e., charge quantum number) in the longer range spin-1 dipole-conserving systems with three and four site terms.
\subsubsection{$t-J_z$ Model}
A simple example is the $t-J_z$ model in one dimension, which appears in the large-$U$ limit of the Hubbard model~\cite{zhang1997tJz, batista2000tJz}.
The model describes the nearest-neighboring hopping of spin-1/2 fermions on a chain, within the constrained Hilbert space that forbids a double occupancy of sites.
Denoting spins by $\sigma \in \{\uparrow, \downarrow\}$ and the corresponding fermion creation and annihilation operators by $\cd_{j, \sigma}$ and $c_{j,\sigma}$, the explicit Hamiltonian in one dimension can be expressed in terms of constrained fermion operators $\tilde{c}_{j,\sigma} \equiv c_{j,\sigma} (1 - \cd_{j,-\sigma} c_{j,-\sigma})$ as~\cite{rakovszky2019statistical}
\begin{equation}
    H_{t-J_z} \equiv \sumal{j}{}{[-t_{j} \sumal{\sigma \in \{\uparrow, \downarrow\}}{}{\left(\tilde{c}_{j,\sigma} \tilde{c}^\dagger_{j+1,\sigma} + h.c.\right)} + J^z_{j} S^z_j S^z_{j+1}]},
\label{eq:tJzhamil}
\end{equation}
where $t_{j}$, $J^z_{j}$ are arbitrary constants, $S^z_j = (\tilde{c}^\dagger_{j, \uparrow}\tilde{c}_{j, \uparrow} - \tilde{c}^\dagger_{j, \downarrow}\tilde{c}_{j, \downarrow})$, and $\tilde{c}_{j,\sigma}$.
In other words, denoting the fermions by $\uparrow$ and $\downarrow$ and vacant sites by $0$, the Hamiltonian only allows the transitions $\ket{\uparrow 0} \leftrightarrow \ket{0 \uparrow}$ and $\ket{\downarrow 0} \leftrightarrow \ket{0 \downarrow}$. 
Given these transitions, it is easy to show that the $t-J_z$ model is fragmented in one dimension.
Since an $\uparrow$ cannot cross over a $\downarrow$ and vice-versa, the full pattern of fermion spins along the chain is conserved~\cite{rakovszky2019statistical}, resulting in exponentially many dynamically disconnected Krylov subspaces that appear within quantum number sectors labelled by the total numbers of $\uparrow$ and $\downarrow$ fermions, the two $U(1)$ symmetries of the system. 
\subsubsection{Miscellaneous Examples}
Other notable examples of fragmentation typically appear in the presence of hard constraints that naturally arise in effective Hamiltonians obtained by a truncation of the Schrieffer-Wolff transformation~\cite{bravyi2011SW} in the presence of a large parameter, e.g. the dipole-conserving models in the presence of a large electric field. 
For example, fragmentation similar to that in the $t-J_z$ model also occurs in the $t-V$ model in the strong coupling regime, which is illustrated in Ref.~\cite{detomasi2019dynamics}.
Other such examples include certain one-dimensional models with strict confinement~\cite{yang2019hilbertspace, chen2021emergent, bastianello2021fragmentation}, where the Hamiltonian restricted to a Krylov subspace was shown to be the integrable XXZ model, models within the Fibonacci Hilbert space of the Rydberg blockade~\cite{mukherjee2021minimal, mukherjee2021constraint}, in the presence of frustration~\cite{lee2020frustrationinduced, hahn2021information} or dipolar interactions~\cite{li2021hilbert}.
In addition, several one-dimensional models introduced in earlier literature in various contexts have been shown to exhibit fragmentation, including the Fredkin~\cite{langlett2021hilbert}, Motzkin~\cite{richter2021anomalous}, Pair-Flip~\cite{moudgalya2021hilbert}, folded XXZ~\cite{zadnik2021foldedxxz1, zadnik2021foldedxxz2, pozsgay2021integrable, bidzhiev2021macroscopic}, and Temperley-Lieb spin chains~\cite{readsaleur2007, moudgalya2021hilbert}.
While most examples of fragmentation are in the product state basis, Ref.~\cite{moudgalya2021hilbert} recently showed that one-dimensional models based on the Temperley-Lieb algebra, including the spin-1 biquadratic spin chains, are fragmented in an entangled basis constructed using spin singlets.
\subsubsection{Higher Dimensional Systems}
Relatively fewer models are known to exhibit fragmentation in dimensions higher than 1.
Several models that are fragmented in one dimension, for example the $t-J_z$ model of Eq.~(\ref{eq:tJzhamil}), are no longer fragmented in higher dimensions~\cite{moudgalya2021hilbert}.
Nevertheless, Ref.~\cite{khemani2019local} argued that the conservation of dipole moment in all directions on hypercubic lattices is sufficient to guarantee the existence of exponentially many frozen eigenstates.
In addition, they showed that the conservation of dipole and quadrupole moments in two dimensions is sufficient for the existence of several other features of fragmentation, including blockades that dynamically disconnect different parts of the system.
Meanwhile Ref.~\cite{khudorozhkov2021hilbert} studied fragmentation in two-dimensional ring-exchange models, which in addition to conserving dipole and quadrupole moments, also possess certain subsystem symmetries. 
\subsubsection{General features of fragmented systems}\label{subsubsec:genfeatures}
We now comment on a few general features that occur in fragmented systems.
However, note that the nature and consequences of fragmentation differ from model to model, and so far there is no universally accepted defining feature of fragmentation (see Ref.~\cite{moudgalya2021hilbert} for a proposed definition).
Three common features of systems exhibiting fragmentation are: (i) Multiple types of Krylov subspaces where all parts of the system are ``active", (ii) Exponentially many product states that are completely frozen, (iii) Frozen regions that lead to blockades in the system that effectively disconnect ``active" regions of the system. 
As discussed in Sec.~\ref{subsec:dipoleconservation}, the dipole-conserving systems possess all three features. 
Other examples of fragmentation in the literature typically possess some of these features, although not necessarily all of them. 
For example, several models discussed in the previous paragraph, including the $t-J_z$ model, do not possess feature (iii), i.e., it is not possible to construct frozen regions that disconnect regions of the system.   
On the other hand, the PXP model discussed in Sec.~\ref{subsec:PXPmodels}, from the point of view of the full spin-1/2 Hilbert space, can possess nearest-neighboring excitations $\ket{\cdots \uparrow \uparrow \cdots}$ that are examples of ``frozen regions" unaffected by the action of the Hamiltonian of Eq.~(\ref{eq:PXPdefn}).
These frozen regions dynamically disconnect different parts of the system, leading to exponentially many Krylov subspaces;\footnote{The experimentally relevant Krylov subspace among these is the Fibonacci Hilbert space, the one without any nearest-neighbor excitations, which is the focus of studies on the PXP model~\cite{SerbynReview}.} hence the PXP model is a trivial example of fragmentation that only possesses feature (iii).
\subsection{Implications to dynamics and connections to QMBS}\label{subsec:fragmentationdynamics}
\begin{figure}
\centering
    \includegraphics[scale=0.25]{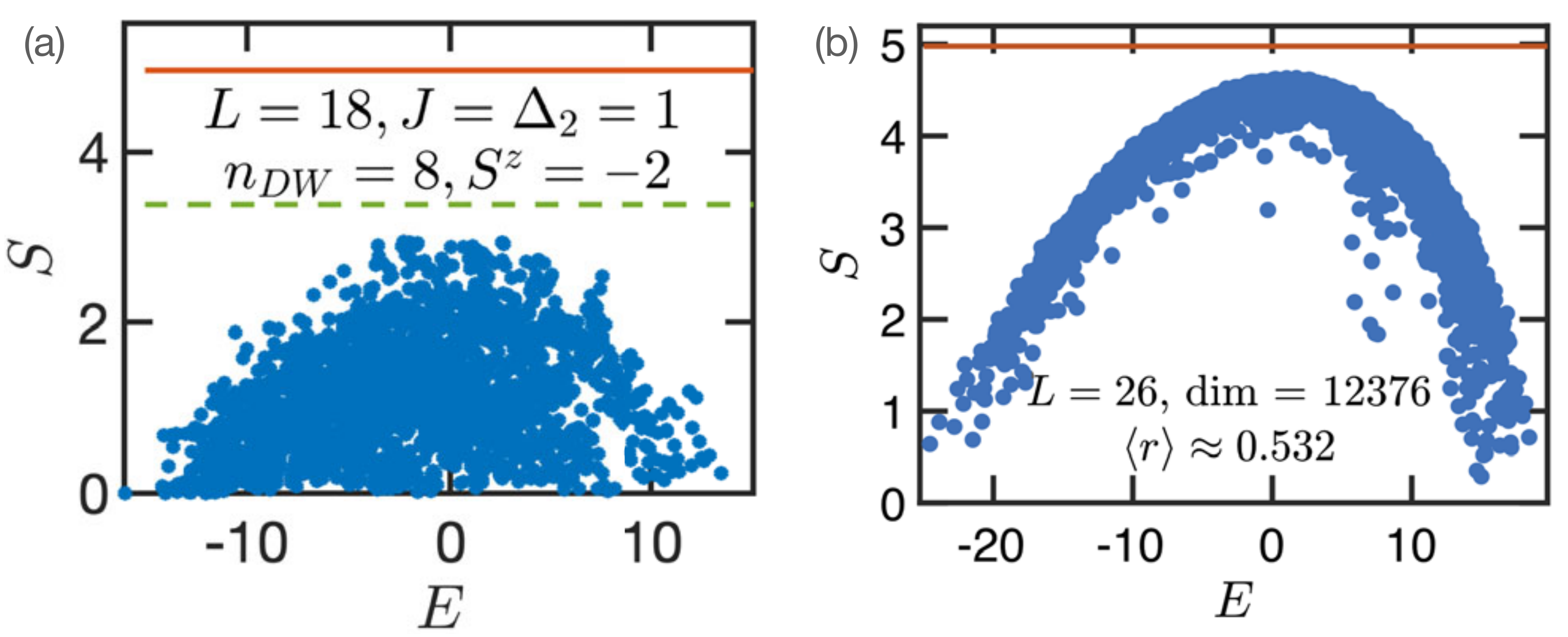}
    \caption{(Color online) Figure reproduced from Ref.~\cite{yang2019hilbertspace} showing the  bipartite entanglement entropy (EE) of eigenstates in a system exhibiting strong Hilbert space fragmentation.
    (a) EE as a function of energy in a particular quantum number sector specified by the total spin $S^z$ and domain wall number $n_{DW}$. 
    The wide distribution of the EE at a particular energy is in stark contrast to EE of eigenstates in typical non-integrable models.  
    (b) EE of eigenstates within a particular Krylov subspace of the fragmented model.
    This plot resembles the behavior of the EE within a quantum number sector of typical non-integrable models. 
    }
    \label{fig:fragmentationproperties_ent}
\end{figure}
From the perspective of the full Hilbert space $\mH$ after resolving quantum numbers of all the conventional symmetries, fragmented systems violate either strong ETH or weak (and hence also strong) ETH~\cite{sala2019ergodicity, khemani2019local}, giving rise to richer physics than QMBS.
\subsubsection{Strong Fragmentation}
\begin{figure}
    \centering
    \includegraphics[scale=0.25]{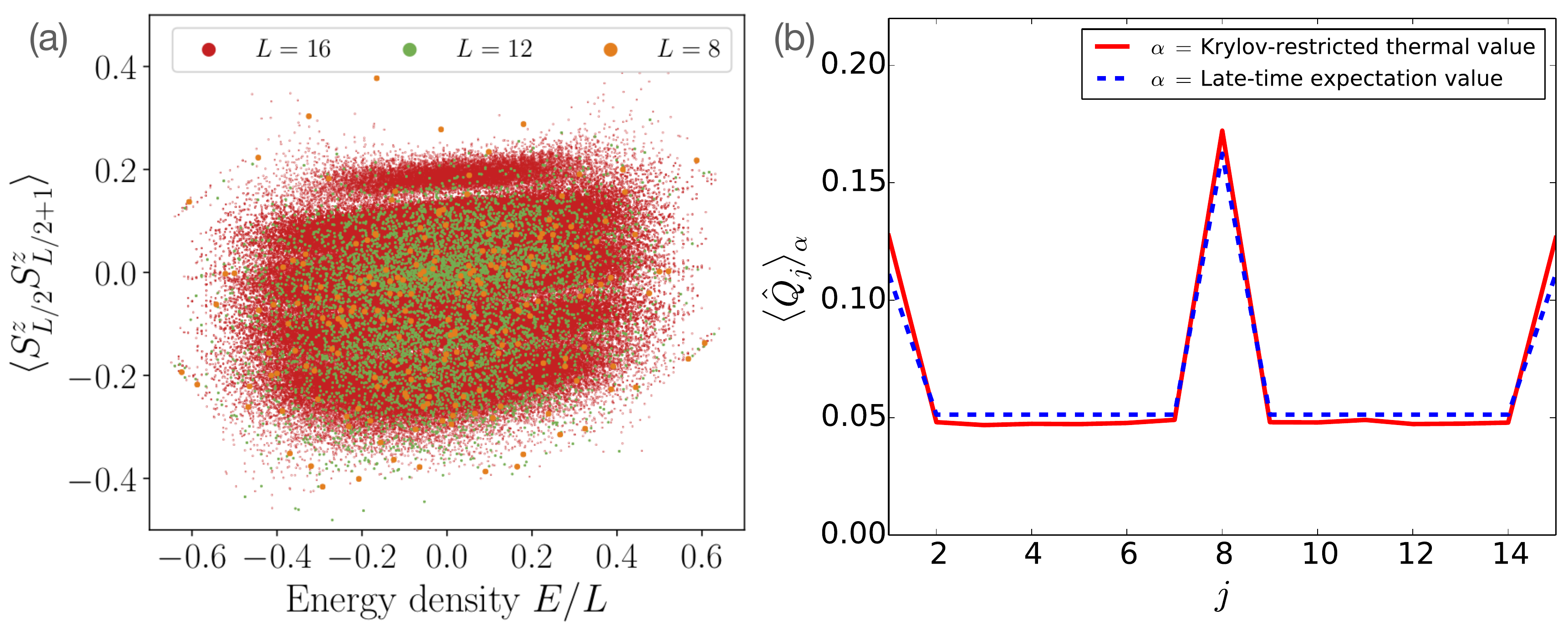}
    \caption{(Color online) ETH in fragmented systems. (a) Figure reproduced from Ref.~\cite{rakovszky2019statistical} showing the expectation value of a certain local operator within eigenstates of the $t-J_z$ model of Eq.~(\ref{eq:tJzhamil}). The large variance in the distribution of the expectation values at a particular energy is in stark contrast to behaviour expected in generic non-integrable models. (b) Figure reproduced from Ref.~\cite{moudgalya2019thermalization} showing the infinite-temperature thermalization of the charge density within a Krylov subspace of the pair-hopping model of Eq.~(\ref{eq:pairhopping}).}
    \label{fig:fragmentationETH}
\end{figure}

Strongly fragmented systems, depicted in Fig.~\ref{fig:main}b, do not have a dominant Krylov subspace in the thermodynamic limit (i.e., $D_{\max}/D \rightarrow 0$ as $L \rightarrow \infty$), and hence violate the conventional form of weak ETH (w.r.t. the full Hilbert space) in addition to strong ETH.
In contrast to examples of QMBS, typical initial states in strongly fragmented systems do not thermalize with respect to the full Hilbert space.
The absence of conventional thermalization can be understood in terms of the EE of eigenstates. 
As discussed in Sec.~\ref{subsec:entanglement}, in the absence of fragmentation, the bipartite EE of a typical eigenstate is given by $S_{\text{th}}$ of Eq.~(\ref{thermalent}) for spin-1/2 systems.
In the presence of fragmentation, the EE of eigenstates that belong to the Krylov subspace $\mK_j$ about a subsystem with $L_{\mA} \leq L/2$ spins is upper bounded by $\sim \log\left({D_{\mK_j}\left[L_{\mA}\right]}\right)$, where $D_{\mK_j}\left[L_{\mA}\right]$ is the dimension of the Krylov subspace $\mK_j$ restricted to the subsystem of size $L_{\mA}$, similar to the EE in constrained systems~\cite{morampudi2020universal, detomasi2020entanglement}.
Similar bounds hold for the EE of a late-time state obtained by time evolution of product states $\ket{\psi_j}$ within a Krylov subspace $\mK_j$. 
In a Krylov subspace whose dimension restricted to a subsystem grows exponentially with subsystem size, this could still result in a volume-law behavior, although with a smaller coefficient.
For example, in a spin-1/2 fragmented system and a Krylov subspace $\mK_j$ where $D_{\mK_j}[L_{\mA}] \sim \phi^{L_{\mA}}$ for $1 < \phi < 2$, the EE is bounded by $S \leq L_{\mA} \log \phi < L_{\mA} \log 2$.
On the other hand, the entanglement entropy for product states that are part of $\mathcal{O}(1)$-dimensional Krylov subspaces cannot exceed a constant value, resulting in a more apparent breakdown of thermalization. 
Moreover, in fragmented systems that exhibit blockades discussed in Sec.~\ref{subsubsec:genfeatures}, randomly chosen product states typically consist of regions that are frozen under the action of the Hamiltonian, which could lead to a further breakdown of thermalization.
For example, the entanglement entropy of all eigenstates within a Krylov subspace with a blockaded region is zero if the bipartition cut is within the blockaded region.
This variety of Krylov subspaces in fragmented models gives rise to a large variance of the EE of eigenstates even within a conventional quantum number sectors, e.g. as shown in Fig.~\ref{fig:fragmentationproperties_ent}a.
However, this variance is small for eigenstates within a Krylov subspace, as shown in Fig.~\ref{fig:fragmentationproperties_ent}b.
The existence of large Krylov subspaces, even after resolving conventional symmetry quantum numbers, also has direct implications for the level statistics and ETH in systems that exhibit strong fragmentation. 
For example, an evident lack of level repulsion is observed between energy levels that belong to different Krylov subspaces, hence systems with strong fragmentation exhibit Poisson level statistics even after resolving conventional symmetry quantum numbers.
In addition, since the eigenstates in different Krylov subspaces are ``uncorrelated", diagonal ETH is generically violated.
In particular, the eigenstate expectation values of local operators are not smooth functions of energy even within a given quantum number sector of conventional symmetry, e.g., as shown for the $t-J_z$ model is shown in Fig.~\ref{fig:fragmentationETH}a.
This should be contrasted with the situation for typical non-integrable models, where the eigenstate expectation values of local operators are given by the thermal value at that energy, up to finite-size corrections (see Eq.~(\ref{eq:ethmatrixel})).
This is also different from models with QMBS, where diagonal ETH is satisfied in almost all eigenstates, apart from a small set of QMBS eigenstates, e.g., as shown in Fig.~\ref{fig:isolatedQMBS}.  
\subsubsection{Weak fragmentation}
Unlike strongly fragmented systems, weakly fragmented ones have a single dominant non-integrable Krylov subspace and its dimension approaches the dimension of the full Hilbert space in the thermodynamic limit (i.e, $D_{\max}/D \rightarrow 1$ as $L \rightarrow \infty$).
Hence, while they violate strong ETH due to frozen eigenstates and $\mathcal{O}(1)$-dimensional Krylov subspaces, they generically satisfy weak ETH as a consequence of the dominant block.
Thus, typical initial states thermalize with respect to the full Hilbert space in weakly fragmented systems, although particular initial states that have large weight on the small Krylov subspaces do not thermalize.
This is also evident from conventional diagnostics such as the energy level statistics and entanglement entropy in weakly fragmented systems, which obey the same behavior as models with QMBS, summarized in Tab.~\ref{tab:taxonomy}.
Indeed, weakly fragmented systems share a lot of their phenomenology with QMBS depicted in Fig.~\ref{fig:main}a, and the exponentially many eigenstates that do not belong to the dominant Krylov subspace should be considered examples of QMBS, according to the definitions discussed in Sec.~\ref{subsec:ergodicitybreaking}.
Most of these eigenstates are generically not equally spaced in energy,  hence they are examples of isolated QMBS discussed in Sec.~\ref{sec:isolated} as opposed to towers of QMBS discussed in Sec.~\ref{sec:towers}.\footnote{Note that as demonstrated in Refs.~\cite{detomasi2019dynamics, hudomal2020quantum, desaules2021proposal}, it is possible to construct equally spaced eigenstates in some fragmented systems, particularly if the system can possess blockades. The existence of such eigenstates leads to revivals from particular initial states, giving rise to phenomenology of towers of QMBS in some fragmented systems. However, this is not generically the case.} 
\subsubsection{Krylov-Restricted Thermalization}
In spite of ETH violation with respect to the full Hilbert space in fragmented systems, expectation values of local operators within eigenstates of sufficiently large (with dimension $\mathcal{D}_{\mathcal{K}}[L] \rightarrow \infty$ as $L \rightarrow \infty$) Krylov subspaces $\mK_j$ do show signatures of ETH, even in strongly fragmented systems.
This phenomenon was referred to as Krylov-Restricted Thermalization in Ref.~\cite{moudgalya2019thermalization}, and evidence for its validity has been found in various fragmented systems~\cite{moudgalya2019thermalization, moudgalya2020quantum, yang2019hilbertspace, hahn2021information}.
In strongly fragmented systems, the Krylov-Restricted Thermalization within the non-locally constrained Krylov subspaces $\mK_j$ can lead to many surprising consequences, including atypical late-time expectation values of local operators~\cite{moudgalya2019thermalization, hahn2021information}, and an apparent Casimir effect~\cite{feng2021hilbert}.
For example, the infinite-temperature charge density profile within a Krylov subspace of the pair-hopping model of Eq.~(\ref{eq:pairhopping}) has been shown in Fig.~\ref{fig:fragmentationETH}b.
While Krylov-Restricted ETH is novel, Refs.~\cite{detomasi2019dynamics, herviou2021manybody} also showed the existence of an ETH-MBL transition within certain non-integrable Krylov subspaces of a spin-1/2 dipole conserving model, which should constitute a novel form of non-locally constrained MBL in fragmented systems that might be different from locally constrained MBL~\cite{chen2018localize, mondragonshem2020fate}.
In addition, Ref.~\cite{bastianello2021fragmentation} found the emergence of ballistic transport and phenomenology associated with integrable systems, in a fragmented model possessing some integrable Krylov subspaces.
\subsubsection{Labelling Krylov Subspaces}
The results on Krylov-restricted thermalization show that large enough Krylov subspaces in fragmented systems closely resemble quantum number sectors corresponding to conventional symmetries, and calls for a characterization of Krylov subspaces in the same language as conventional symmetries. 
Such a question was first explored in Ref.~\cite{rakovszky2019statistical}, where operators referred to as ``Statistically Localized Integrals of Motion" (SLIOMs) were introduced for the $t-J_z$ model and the minimal-range spin-1 dipole moment conserving model with OBC.
Remarkably, the full set of eigenvalues under all the SLIOMs uniquely label the Krylov subspaces.
However, unlike operators corresponding to conventional symmetries, these SLIOMs are highly non-local operators, although they are ``localized" in a sense defined in Ref.~\cite{rakovszky2019statistical} and can be used to explain the anomalous late-time behavior of autocorrelation functions~\cite{rakovszky2019statistical, moudgalya2021hilbert}.
In particular, for certain models of fragmentation, the existence of SLIOMs implies boundary localization and analogues of strong zero modes~\cite{fendley2012parafermionic, fendley2016strong, alicea2016topological} for non-integrable models.
Apart from the $t-J_z$ model, the SLIOMs have also proved useful in explaining dynamical phenomena in other models of fragmentation, e.g., in the context of dimer models~\cite{feldmeier2021emergent}.
While it is not clear if SLIOMs can be constructed in all examples of fragmentation, more recently, Ref.~\cite{moudgalya2021hilbert} approached the question of labelling Krylov subspaces using the language of so-called ``commutant algebras", which generalizes the notion of the symmetry algebra for systems with conventional symmetries.
In particular, the commutant algebra is defined for a family of systems (e.g. for $\{\sum_j{J_j H_j}\}$ in the case of the pair-hopping model of Eq.~(\ref{eq:pairhopping}), and it is the algebra of operators that commute with that family.
Equivalently, it is the algebra of all operators that commute with each term of the Hamiltonian, e.g. with all the $H_j$ in the pair-hopping model. 
Ref.~\cite{moudgalya2021hilbert} showed that the irreducible representations of this commutant algebra uniquely define the Krylov subspaces corresponding to a family of systems.
Further, the algebra also contains all the local and non-local conserved quantities of that family of systems, including the SLIOMs, which uniquely label the Krylov subspaces.
Ref.~\cite{moudgalya2021hilbert} full commutant algebras and the corresponding Krylov subspaces were explicitly constructed for several fragmented models, including systems where the definition of SLIOMs is not straightforward.
Finally, we note that fragmentation in certain models occurs due to the presence of strictly localized integrals of motion, i.e., operators with support on a small number of consecutive sites.
For example, in the PXP model discussed in Sec.~\ref{subsubsec:genfeatures}, the projector $\ket{\uparrow \uparrow}\bra{\uparrow\uparrow}$ onto a nearest-neighbor configuration with excited Rydberg atoms is a strictly local conserved quantity. 
Such conserved quantities are analogous to Local Integrals of Motion (LIOMs) that occur in systems exhibiting many-body localization~\cite{nandkishore2015many}.
Such examples were recently referred to as ``local fragmentation" in Ref.~\cite{buca2021local}, 
and it is then straightforward to label the Krylov subspaces using the LIOMs.
This systems should be contrasted with systems such as dipole conserving models, that exhibit ``true fragmentation", where it is not clear if strictly local conserved quantities exist. %
\subsection{Autocorrelation Functions}
\begin{figure}
    \centering
    \includegraphics[scale=1]{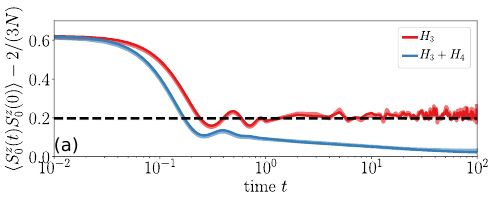}
    \caption{(Color online) Figure reproduced from Ref.~\cite{sala2019ergodicity} showing the  autocorrelation function of the local spin operator in a spin-1 dipole-moment-conserving system exhibiting strong (red) and weak (blue) fragmentation for several system sizes $N$. 
    While autocorrelation functions of local operators in generic non-integrable systems are expected to decay to zero at large times for system sizes , the strongly fragmented dipole-conserving Hamiltonian $H_3$ of Eq.~(\ref{eq:H3hamil}).
    }
    \label{fig:fragmentationproperties_auto}
\end{figure}
Similar to QMBS, some of the key diagnostics used in the literature to identify Hilbert space fragmentation are the bipartite EE, the energy level statistics, and the expectation values of local operators in eigenstates to test ETH, diagnostics discussed in Sec.~\ref{sec:ergodicity}.
In addition to these conventional measures, operator spreading (or lack thereof) has played a key role in diagnosing and understanding fragmented systems~\cite{pai2018localization}. 
In particular, the spreading of a local operator can be characterized by its autocorrelation function, and we discuss its properties below. 
\subsubsection{Definition and Properties}
The infinite-temperature autocorrelation function of a local operator $\hO$ under the Hamiltonian $H$ is defined as
\begin{equation}
    \mA_L(t) \equiv \langle \hO(t) \hO(0)\rangle = \frac{1}{D}\textrm{Tr}{(e^{i H t} \hO e^{-i H t} \hO)},
\label{eq:autocorr}
\end{equation}
where $D$ is the Hilbert space dimension and $L$ is the system size. 
The behavior of autocorrelation functions of local operators, both as a function of time and system size, contain a lot of information about the system, including its symmetries and the nature of transport.
In the infinite-time limit, the saturation value of the autocorrelation function, or more precisely the infinite-time average defined as $\bar{\mA}_L \equiv \lim_{T \rightarrow \infty}{\frac{1}{T}\int_0^T{dt\ \mA_L(t)}}$, is an indicator of the symmetries of the Hamiltonian $H$.  
Indeed, the Mazur bound that provides a lower bound on $\bar{\mA}_L$ is stated in terms of the overlap of the operator $\hO$ on the various conserved quantities of the system~\cite{mazurbound1969,suzukiequality1971,dhar2020revisiting, rakovszky2019statistical,moudgalya2021hilbert}, and this bound is generically saturated. 
For example, in a system with $U(1)$ charge conservation, $\bar{\mA}_L$ is expected to decay with system size as $\bar{\mA}_L \sim 1/L$, a scaling that can be understood in terms of Mazur bounds~\cite{rakovszky2019statistical, moudgalya2021hilbert}.
In the infinite-size limit, where the quantity of interest is $\mA_\infty(t) \equiv \lim_{L \rightarrow \infty}\mA_L(t)$, the decay of the autocorrelation function with time reveals the nature of transport in the system.
For example, in diffusive systems, this autocorrelation function in the thermodynamic limit is expected to decays with time $\mA_{\infty}(t) \sim 1/\sqrt{t}$, which also forms a numerical diagnostic of the nature of transport in the system. 
Similarly, the system is said to be subdiffusive if $\mA_{\infty}(t) \sim 1/t^\beta$ for $\beta < 1/2$, which is observed in several constrained systems, including ones that conserve dipole moment~\cite{feldmeier2020anomalous, morningstar2020kinetically, iaconis2019anomalous, iaconis2021multipole, moudgalya2021spectral}.
Below we discuss some of aspects of these diagnostics in fragmented systems.
\subsubsection{Strong Fragmentation}
Strong fragmentation is also usually associated with an anomalous saturation of the infinite-time autocorrelation function $\bar{\mA}_L$ of local operators as a function of system size. 
For example, strongly fragmented dipole-moment conserving systems, e.g., the pair-hopping model of Eq.~(\ref{eq:pairhopping}) or the spin-1 model of Eq.~(\ref{eq:H3hamil}), are known to exhibit ``operator localization"~\cite{pai2018localization}.
That is, when $\hO$ is chosen to be a strictly local operator on a site $j$, the weight the time-evolved operator on the original site $j$ is finite even at infinite times, which is reflected in the finite value of $\bar{\mA}_L$ even in the the thermodynamic limit, e.g., as shown in the red curve in Fig.~\ref{fig:fragmentationproperties_auto}.
This is contrast to systems with conventional symmetries, where local operators are expected to spread uniformly throughout the system at late times, and $\bar{\mA}_L$ is expected to decay with system size.  
This operator localization in dipole-conserving systems can be understood to be a consequence of the abundance of Krylov subspaces with blockades~\cite{rakovszky2019statistical,moudgalya2021hilbert}, closely related to strong fragmentation.
Similar effects are also observed in the autocorrelation functions of local operators in the $t-J_z$ model~\cite{rakovszky2019statistical, moudgalya2021hilbert}, where certain operators on the boundary are localized. 
In addition, the autocorrelation functions of local operators in the bulk exhibit a decay of $\bar{\mA}_L \sim 1/\sqrt{L}$, whereas a decay of $\bar{\mA}_L \sim 1/L$ is expected from conventional symmetry considerations.
\subsubsection{Weak fragmentation}

In addition, unlike strongly fragmented systems, the effect of weak fragmentation can safely be ignored in the context of infinite-temperature properties of the system, e.g., in the autocorrelation functions $\mA_L(t)$ of local operators. 
This is the case in dipole-moment conserving systems, for example, where there has been a large interest in its sub-diffusive transport at infinite-temperature~\cite{feldmeier2020anomalous, guardado2020subdiffusion, zhang2020subdiffusion, moudgalya2021spectral}.
Even though dipole-moment conserving systems with finite-range terms are always fragmented, the fragmentation is weak beyond a certain range (e.g., in the pair-hopping model discussed in Sec.~\ref{subsec:dipoleconservation}) and the autocorrelation functions of local operators decay as $\bar{\mA}_\infty(t) \sim 1/t^{\frac{1}{4}}$, which can be understood without taking fragmentation into consideration. 
Furthermore, the infinite-time autocorrelation function $\bar{\mA}_L$ also decays with system-size, e.g., as shown in the blue curve in Fig.~\ref{fig:fragmentationproperties_auto}, indicating the absence of operator localization. 
This distinction in the behavior of autocorrelation functions (as well as out-of-time-ordered correlation functions (OTOCs)) between strongly and weakly fragmentated systems has also been used to study the ``freezing transition" from weak to strong fragmentation~\cite{morningstar2020kinetically, feldmeier2021critically}.
\section{Discussion and Outlook}\label{sec:outlook}
\begin{figure}
    \centering
    \includegraphics[scale=0.9]{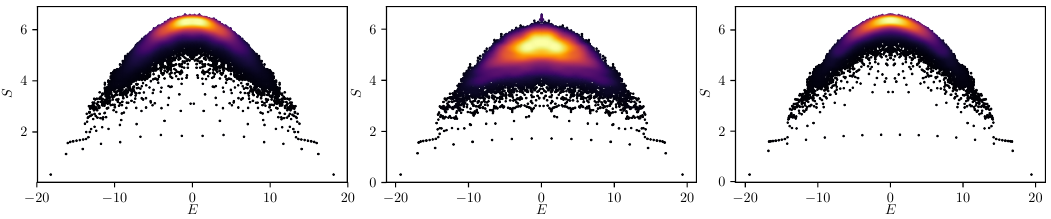}
    \caption{(Color online) Figure reproduced from Ref.~\cite{choi2018emergent} showing the ``enhancement" of the QMBS in the PXP model with the addition of a particular perturbation. This result, among many others (e.g., those discussed in Refs.~\cite{lin2019exact, iadecola2019quantum}), hints towards a connection between examples of exact towers of QMBS discussed in this review and the approximate QMBS in the PXP model.
}
    \label{fig:outlook_pxp}
\end{figure}
Despite being a relatively new field, QMBS has already attracted a large attention as exemplified in this review.
Nevertheless, several major open questions remain, and we summarize a few of those in the following.
\subsubsection*{Unified framework for QMBS}
An immediate question is the development of a unified language to describe and understand known examples of QMBS, which might lead to a finer classifications of QMBS.
As discussed in Sec.~\ref{sec:unifiedformalism}, some progress in this direction has been made with the introduction of embedding, SGA-based, and symmetry-based formalisms to explain several examples of towers of QMBS.
However, the precise relations between the various formalisms are not yet clear.
Several examples of towers of QMBS discussed in Sec.~\ref{subsec:othertowers}, including the QMBS in the spin-1 AKLT model, fall outside the Shiraishi-Mori (SM) embedding and symmetry-based formalisms.
On the other hand, unlike the embedding and symmetry-based formalisms, the SGA-based formalisms lack a precise prescription for constructing models with QMBS. 
Furthermore, many examples have not been explicitly shown to be captured by any of the formalisms, for example the second tower of QMBS in the spin-1 XY model~\cite{schecter2019weak, chattopadhyay2019quantum}, which are not expressed as a repeated action of a raising operator on a simple eigenstate.
For isolated examples of QMBS discussed in Sec.~\ref{sec:isolated}, the SM formalism captures several examples in the literature, although the connection is not always immediately apparent.
Yet there are examples of isolated QMBS such as the exact eigenstates in 2d and higher-dimensional PXP models discussed in Sec.~\ref{subsec:PXPmodels} that are yet to be understood in this approach. 
A better understanding of the QMBS formalisms will also help extend examples of QMBS, many of which are restricted to one-dimensional systems, to higher dimensions.
Moreover, higher dimensions might also reveal qualitatively different types of QMBS, such as the exponentially many QMBS in the 2D PXP model discussed in Sec.~\ref{subsec:PXPmodels}, which have no 1D counterparts. 
\subsubsection*{Stability of QMBS}
Another important question that requires further investigation is the stability of exact QMBS to perturbations.
One aspect of stability is whether the QMBS eigenstates survive perturbations in the thermodynamic limit. 
Refs.~\cite{lin2020slow} and \cite{surace2021exactstability} explored this question for the exact QMBS in the 1d PXP model~\cite{lin2019exact} and the tower of QMBS in the spin-1 XY model~\cite{schecter2019weak}, and found evidence that in the thermodynamic limit, QMBS are unstable to generic perturbations, i.e., they hybridize with thermal eigenstates for arbitrary small perturbation strengths.
Nevertheless, the thermalization times for local observables in the perturbed model was found to be finite and parametrically large (i.e., diverging as the perturbation strength goes to zero) even in the thermodynamic limit~\cite{lin2020slow}, and the QMBS eigenstates displayed anomalous robustness at first order perturbation theory~\cite{surace2021exactstability}.
These results show that the exact QMBS in the PXP model do have some degree of stability under perturbations even in the thermodynamic limit, and it would be interesting to systematically probe this question for other examples of QMBS, particularly for exact towers discussed in Sec.~\ref{sec:towers}.
Irrespective of their stability in the thermodynamic limit, a more experimentally relevant question in the current era of quantum simulators and Noisy Intermediate Scale Quantum (NISQ) devices~\cite{preskill2018quantumcomputing} is the stability of QMBS signatures in finite-size systems.
For example, we can ask whether signatures of QMBS such as anomalous dynamics on unexpectedly long time-scales persist under perturbations, or if for finite system sizes, approximate QMBS survive in models that are proximate to those with exact QMBS. 
A classic example is the PXP model, where approximate QMBS seem rather robust to perturbations~\cite{choi2018emergent, mondragonshem2020fate, turner2021correspondence, bull2020quantum, surace2021quantum}, and experimentally show slowly-decaying revivals~\cite{bernien2017probing, bluvstein2020controlling}.
These experimental setups consist of 51-200 Rydberg atoms, far from the thermodynamic limit, which motivates the study of the stability of QMBS at finite system-sizes. 
The QMBS in the PXP model have been studied using a wide variety of techniques that yield several insights into the origin of the approximate QMBS~\cite{SerbynReview, ho2018periodic, michailidis2020slow, turner2021correspondence, bull2020quantum}, and several phenomenological results are known about the PXP models, and their deformations~\cite{khemani2019int, choi2018emergent, bull2019systematic, yao2021quantum}. 
Nevertheless, a major open question in this field is to precisely connect these results on the approximate QMBS in the PXP to exact QMBS in various other systems.
In particular, can these approximate QMBS be understood due to its proximity to a model with exact QMBS?
Evidence supporting this was shown in Refs.~\cite{lin2019exact, mark2019new, iadecola2019quantum}, which found approximate momentum $\pi$ multi-quasiparticle descriptions of the PXP QMBS.
Moreover, Ref.~\cite{khemani2019int} found a Hamiltonian proximate to the PXP model that shows an atypical behavior of level statistics with system size, which they conjectured to be an integrable point.
Later, Ref.~\cite{choi2018emergent} found a different Hamiltonian proximate to the PXP model that exhibits almost perfect revivals and a greatly enhanced decoupling of the QMBS subspace from the rest of the spectrum (see Fig.~\ref{fig:outlook_pxp}).
These works reveal that the phenomenology of the approximate QMBS in the PXP model resembles that of exact towers of QMBS discussed in Sec.~\ref{sec:towers}, and suggest that exact QMBS might have more stability than currently believed, at least for present-day experimentally accessible system-sizes.

\subsubsection*{Floquet QMBS}
The exploration of QMBS beyond Hamiltonian systems, for example in Floquet systems, is also an interesting direction of study.
Exact QMBS in the PXP model~\cite{lin2019exact} were extended to Floquet-PXP Hamiltonians in Ref.~\cite{mizuta2020exact}, and some of them were shown to be intrinsic to Floquet systems, arising only at particular drive frequencies~\cite{sugiura2021manybody}.
A related result is the construction of exact eigenstates based on short orbits in a cellular automaton that is obtained in an appropriate limit of the Floquet-PXP Hamiltonian in Ref.~\cite{iadecola2020nonergodic}. 
Moreover, it should be possible to generalize some of unified formalisms, particularly the SM formalism to Floquet systems, which might lead to Floquet analogues of some of the QMBS discussed in this review. 
The exploration of QMBS in Floquet systems is particularly interesting since recent experimental and numerical results suggest that QMBS in the PXP model can be stabilized under periodic driving~\cite{bluvstein2020controlling, maskara2021discrete}. 
Furthermore, obtaining more analytically tractable examples might shed light on the many numerical results on QMBS in driven systems~\cite{pai2019dynamical, mukherjee2020collapse, pizzi2020timecrystal, yarloo2020homogeneous, zhao2020quantum, haldar2021dynamical}. 
\subsubsection*{Fragmentation and symmetries}
\begin{figure}
    \includegraphics[scale=0.6]{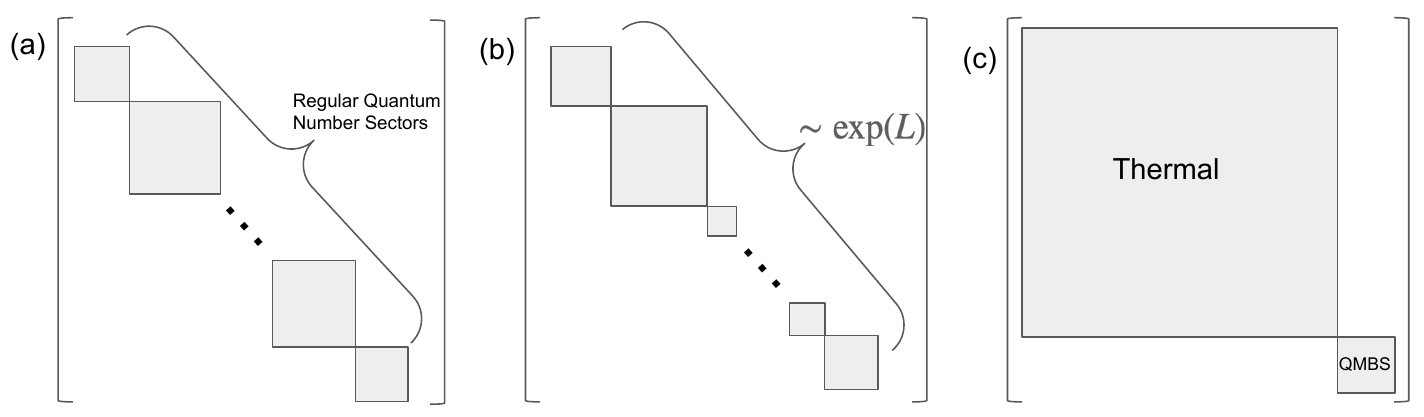}
    \caption{Schematic depiction of the block diagonal structure of the Hamiltonian showing the dynamically disconnected ``Krylov subspaces" in systems with (a)  conventional symmetries  (b) Hilbert space fragmentation (c) QMBS.}
    \label{fig:outlook_QMBS_fragmentation}
\end{figure}
Turning to Hilbert space fragmentation, several open questions are still looming. 
Firstly, the phenomenon of Hilbert space fragmentation clearly demonstrates the need to sharpen the definition of ``symmetry" in a quantum many-body system~\cite{moudgalya2021hilbert}.
The Krylov subspaces that occur in fragmented systems are strongly reminiscent of quantum number sectors of conventional symmetry, e.g., as shown in Fig.~\ref{fig:outlook_QMBS_fragmentation}.
Conserved quantities associated with conventional symmetries are typically sums of local terms, or products of on-site unitary operators, and the dynamically disconnected subspaces in such systems are the different eigenspaces of such quantities.
However, in fragmented systems, the Krylov subspaces are labeled by the eigenvalues of certain non-local conserved quantities~\cite{rakovszky2019statistical, moudgalya2021hilbert}, which could be considered as ``non-local symmetries"; however, unlike non-local symmetries that appear in the literature (e.g., in the context of topological systems), these do not have obvious on-site actions.
If arbitrary non-local operators are considered valid conserved quantities, any finite-dimensional Hamiltonian trivially has exponentially many conserved quantities -- the eigenstate projectors; hence the necessity of a better definition, or a more practical and experimentally motivated one.
This might also help settle debates~\cite{mondaini2018comment, shiraishimorireply} about which symmetries/Krylov subspaces are needed to be resolved in order to test ETH. 
Understanding the precise nature of ETH-violation in fragmented systems is important, since restricted versions of ETH and MBL have been found to hold within sufficiently large Krylov subspaces, leading to notions of Krylov-Restricted Thermalization and its breakdown.  
\subsubsection*{Analytical examples of fragmentation}
A different direction that needs to be pursued is the search for additional analytically tractable examples of fragmentation, which might also help better understand the necessary and sufficient conditions for fragmentation to occur. 
As discussed in Sec.~\ref{subsec:dipoleconservation}, an experimentally relevant example that is well understood is the case with dipole moment conservation in one dimension.
While a general characterization of all the Krylov subspaces was obtained in the minimal-range spin-1 dipole-conserving model in Ref.~\cite{moudgalya2021hilbert}, a more general understanding of the longer-range dipole-conserving model apart from the minimal-range ones~\cite{moudgalya2019thermalization, rakovszky2019statistical, moudgalya2021hilbert} is lacking, and many of the results, such as the nature of fragmentation (strong or weak), rely on numerical observations.
The exploration of fragmentation in higher dimensions is also important, particularly since two-dimensional systems are sometimes easier to simulate using optical lattices.
Multipole moment conservation laws can be imposed by subjecting systems to particular electric potentials, and while Ref.~\cite{khemani2019local} showed that these are sufficient to guarantee fragmentation, several questions, such as the structure of larger Krylov subspaces, have not been explored.
Hilbert space fragmentation might also be related to several earlier examples of ergodicity breaking due to the formation of dynamical subsectors, such as dynamical localization in gauge theories~\cite{smith2018dynamical}, or localization due to superselection sectors~\cite{kim2016localization}, and it would be interesting to make the connections more precise.
On a different note, several aspects of fragmentation might be relevant in the study of equilibrium physics close to the ground state.
For example, as a consequence of fragmentation, the ground state and some low-lying excited states of some dipole-conserving models that appear in the thin-torus limit of fractional Quantum Hall systems were found to have simple expressions in terms of fragmented Matrix Product States~\cite{nachtergaele2020lowcomplexity, nachtergaele2021spectral}, which proved useful in addressing questions on the gap of such systems~\cite{warzel2021bulk, warzel2021spectral}.

\subsubsection*{Classical versus Quantum fragmentation}
Finally, most examples of Hilbert space fragmentation consist of Hamiltonians that are fragmented in the product state basis (``classical fragmentation"), and the Krylov subspaces are completely determined by the transitions between product states allowed by terms of the Hamiltonian, which is essentially a classical process.
The possibility of fragmentation in a more entangled basis (``quantum fragmentation") was recently pointed out in the spin-1 biquadratic model~\cite{moudgalya2021hilbert}, but the dynamics in such systems is relatively unexplored, and it remains to be understood whether such fragmentation leads to qualitatively new dynamical phenomena absent in simpler models.
Exploring fragmentation and Krylov subspaces in different bases might also help distinguish between or establish a relation between QMBS discussed in Secs.~\ref{sec:towers} and \ref{sec:unifiedformalism} and the phenomenon of weak Hilbert space fragmentation, which share several common features.
In particular, can towers of QMBS be understood in the same language as Hilbert space fragmentation, since the subspace spanned by the QMBS can be viewed as an small Krylov subspace within the full Hilbert space~\cite{bull2020quantum}  (see Fig.~\ref{fig:outlook_QMBS_fragmentation})?
QMBS and Hilbert space fragmentation have already attracted a large attention thanks to their experimental realization in quantum simulators and the existence of an abundance of exact results and toy models, a rarity in the realm of strongly correlated quantum systems.
However the number of open challenges that we have tried to expose here will undoubtedly be a source of rich discussions and physics, and guarantees a bright and exciting future for this field.
{\it Note added} --- While this review was in preparation, Ref.~\cite{papic2021weak} appeared, which provides a complementary pedagogical review of aspects of quantum many-body scars and Hilbert space fragmentation.
\ack
We are particularly grateful to Lesik Motrunich for enlightening discussions. 
We also acknowledge useful discussions with Berislav Buca, Dumitru Calugaru,  Paul Fendley, David Huse, Tom Iadecola, Frank Pollmann, and Pablo Sala.
We thank Stephan Rachel, Abhinav Prem, Rahul Nandkishore, Ana Hudomal, Ivana Vasic, Zlatko Papic, Loic Herviou, Jens Bardarson, Edward O'Brien, Paul Fendley, and Lesik Motrunich for previous collaborations on related topics.
We are grateful to Yichen Huang, Tom Iadecola, Igor Klebanov, Kiryl Pakrouski, Zlatko Papic, Fedor Popov, Frank Pollmann, Lesik Motrunich, Pablo Sala, and Lenart Zadnik for their comments and feedback on an earlier version of this review.
This work is part of a project that has received funding from the European Research Council (ERC) under the European Union's Horizon 2020 research and innovation programme (grant agreement no. 101020833).
This work is supported by the Walter Burke Institute for Theoretical Physics at Caltech and the Institute for Quantum Information and Matter.
S.M. acknowledges the hospitality of the Aspen Center for Physics, where a part of this work was completed. 
The Aspen Center for Physics is supported by National Science Foundation grant PHY-1607611.
This work was also partially supported by a grant from the Simons Foundation.
\section*{References}
\bibliographystyle{iopart-num-mod}
\bibliography{review}

\end{document}